\documentclass[%
 aip,
 nofootinbib,
 amsmath,amssymb,
 reprint,%
]{revtex4-1}

\usepackage{graphicx}% Include figure files
\usepackage{dcolumn}% Align table columns on decimal point
\usepackage{bm}% bold math
\usepackage[utf8]{inputenc}
\usepackage[T1]{fontenc}
\usepackage{mathptmx}
\usepackage{subfig}
\usepackage{multirow}
\usepackage{soul,color}

\begin{document}

\title[]{Predictive modeling approaches in laser-based material processing}
% Force line breaks with \\

\author{Maria-Christina Velli}
%\altaffiliation[Also at ]{Physics Department, XYZ University.}
%\homepage{http://www.Second.institution.edu/~Charlie.Author.}

\affiliation{ 
Institute of Electronic Structure and Laser, Foundation for Research and Technology - Hellas, N. Plastira 100, Vassilika Vouton, 70013, Heraklion, Greece}%
\affiliation{ 
Department of Physics, University of Crete, P.O. Box 2208, 71003, Heraklion, Greece}

\author{George D. Tsibidis}
\email{tsibidis@iesl.forth.gr}
\affiliation{ 
Institute of Electronic Structure and Laser, Foundation for Research and Technology - Hellas, N. Plastira 100, Vassilika Vouton, 70013, Heraklion, Greece}

\author{Alexandros Mimidis}

\affiliation{ 
Institute of Electronic Structure and Laser, Foundation for Research and Technology - Hellas, N. Plastira 100, Vassilika Vouton, 70013, Heraklion, Greece}
\affiliation{ 
Department of Material Science, University of Crete, P.O. Box 2208, 71003, Heraklion, Greece}

\author{Evangelos Skoulas}

\affiliation{ 
Institute of Electronic Structure and Laser, Foundation for Research and Technology - Hellas, N. Plastira 100, Vassilika Vouton, 70013, Heraklion, Greece}
\affiliation{ 
Department of Material Science, University of Crete, P.O. Box 2208, 71003, Heraklion, Greece}

\author{Yannis Pantazis}

\email{pantazis@iacm.forth.gr}
\affiliation{Institute of Applied and Computational Mathematics, Foundation for Research and Technology - Hellas, N. Plastira 100, Vassilika Vouton, 70013, Heraklion, Greece}

\author{Emmanuel  Stratakis}

\email{stratak@iesl.forth.gr}

\affiliation{ 
Institute of Electronic Structure and Laser, Foundation for Research and Technology - Hellas, N. Plastira 100, Vassilika Vouton, 70013, Heraklion, Greece}%
\affiliation{ 
Department of Physics, University of Crete, P.O. Box 2208, 71003, Heraklion, Greece}

%\affiliation[*]{Corresponding authors: tsibidis@iesl.forth.gr, pantazis@iacm.forth.gr, stratak@iesl.forth.gr}

\date{\today}% It is always \today, today,

\begin{abstract}

Predictive modelling represents an emerging field that combines existing and novel methodologies aimed to rapidly understand physical mechanisms and concurrently develop new materials, processes and structures. In the current study, previously-unexplored predictive modelling in a key-enabled technology, the laser-based manufacturing, aims to automate and forecast the effect of laser processing on material structures. The focus is centred on the performance of representative statistical and machine learning algorithms in predicting the outcome of laser processing on a range of materials. Results on experimental data showed that predictive models were able to satisfactorily learn the mapping between the laser’s input variables and the observed material structure. These results are further integrated with simulation data aiming to elucidate the multiscale physical processes upon laser-material interaction. As a consequence, we augmented the adjusted simulated data to the experimental and substantially improved the predictive performance, due to the availability of increased number of sampling points. In parallel, a metric to identify and quantify the regions with high predictive uncertainty, is presented, revealing that high uncertainty occurs around the transition boundaries. Our results can set the basis for a systematic methodology towards reducing material design, testing and production cost via the replacement of expensive trial-and-error based manufacturing procedure with a precise pre-fabrication predictive tool. 

\end{abstract}

\maketitle

\section{Introduction}
\label{intro:sec}

There is an increasing demand to support materials research in the development of novel or improved applications with advanced strategies. Unfortunately, nowadays, manufacturing processes in a vast range of applications in areas such as automotive, aerospace, microengineering, telecommunications, biotechnologies, microfluidics, photovoltaics, is still performed using expensive trial and error approaches \cite{chitale2013product}. Thus, conventional manufacturing strategies are expected to lead to financial risks and inhibit competitiveness. Although, technological advances and software engineering have been extensively used by manufacturing companies to provide predictive tools in various fields of engineering (i.e. aerospace \cite{dormohammdi2017damage}, automotive \cite{del2010automotive}, etc.), such instruments for advanced processing of materials has not yet been developed.

A very promising and high-resolution material machining process is performed via using lasers, which are proving to be ideal tools for controlling the energy deposition and respective modifications on the surface, or volume, of a material. In particular, material processing  with  femtosecond  (fs) pulsed  lasers has  received considerable  attention  due  to the fact that it is related to a high precision, rapid energy delivery and minimisation of the heat affected area.  Direct  fs-laser  surface micro-and  nano-patterning has  been  demonstrated  in  many  types  of  materials  including semiconductors,  metals,  dielectrics,  ceramics,  and  polymers. Therefore, a fs-based technology is used to an abundance of diverse applications ranging from micro-device fabrication to optoelectronics, microfluidics  and  biomedicine \cite{vorobyev2013direct,zorba2008biomimetic,bauerle2013laser,papadopoulou2010silicon,simitzi2015laser,papadopoulos2019biomimetic,stratakis2012nanomaterials, stratakis2020_review}. These  applications require a thorough knowledge of the fundamentals of laser interaction with the target material for enhanced controllability of the resulting modification of the target relief. Physical mechanisms that lead to surface modification have been extensively explored both theoretically and experimentally \cite{tsibidis2012dynamics,tsibidis2015ripples,tsibidis2014controlled,derrien2013possible,skolski2012laser,bonse2012femtosecond,garrelie2011evidence,wang2005ultrafast,huang2009origin,bonse2010pulse,tsibidis2016convection,shimotsuma2003self,chichkov1996femtosecond,tsibidis2017ripple,tsibidis2020SR}. 

Materials modelling has been a powerful tool that provides key information for tailoring and designing materials or even identifying new materials, providing a cost-effective method and minimising the use of trial and error approaches aiming to reduce the need for an increasing number of experiments. The use of materials modelling in industries is very versatile and it can offer a solution towards controlling the output procedure through a systematic exploration of the, usually, complex physical (multiscale) processes that occur during the manufacturing  processes that occur during the manufacturing process 
\cite{tsibidis2012dynamics,tsibidis2015ripples,tsibidis2017ripple,rethfeld2017modelling,ivanov2010nanocrystalline,lin2008electron}. In principle, the optimum fabrication conditions can be identified via the computational execution of self-consistent virtual experiments. Nevertheless, despite the significance originated from the use of materials modelling, the complexities due to the need of a large number of simulated experiments in which multiscale physical models are involved downgrade the benefit of the methodology as it leads to slow decision making. Therefore, towards improving the decision making process and reduce the production time, further tools are required, based on machine learning-based methods.

A number of predictive modelling approaches based on Machine Learning (ML) techniques were used  for material processing. In Raccuglia et al. \cite{raccuglia2016machine}, an ML model was trained on datasets of `failed' experiments (data that were archived in notebooks stemming from unsuccessful experiments) to predict reaction outcomes for the crystallization of templated Vanadium selenites. Oliynyk et al. \cite{oliynyk2016high} used ML models to study Heusler compounds and properties. In another study, ML methodology was used in laser-based manufacturing to improve geometric accuracy of the fabricated parts \cite{francis2019deep} . On the other hand, an increase of the accuracy and need for prediction of distortion quantification in the fabricated materials was further facilitated with the employment of a Deep Learning approach. Moreover, Tani and Kobayashi \cite{tani2018pulse} performed a big-data analysis to describe how surface morphology affects the laser ablation process. More specifically, a comparison of  a produced  3D depth profile before and after single-shot ablation from thousands of data for various materials, they observed and modeled hysteresis behavior. In another work by Mills et al. \cite{mills2018predictive,heath2018machine}, used a neural network-based approach  to explore the morphology features of an induced 3D surface profile of a substrate after being laser machined with a single laser pulse, for random laser spatial intensity profiles. On the other hand, Agrawal et al.\cite{agrawal2014exploration}   predicted the fatigue strength of steels. Both physics-based and data-driven approaches were used  to correlate properties of alloys and manufacturing process parameters. In that study, data-driven models through extrapolation were able to sample extreme value properties, where the current state-of-the-art physics-based models suffer from severe limitations .

Inspired by the above studies and challenges, we propose to complement laser manufacturing processing and discovery with data-driven analysis. Existing corpora of experimental and simulated measurements constitute a valuable collection of information which remains mostly unexploited when new materials are investigated. Predictive modelling through the utilization of statistical and machine learning models offers an efficient approach to encode the accumulated experience and knowledge from previous experiments into a mathematical model and, subsequently, be able to extrapolate into unexplored conditions. ML models are trainable parametric models that aim to perform a learning task such as classification or regression \cite{DudaHartStork01,bishop2007}. An ML model can be abstractly regarded as a function or a mapping that transforms the input to an output. In order to accurately learn the mapping between the input variables (e.g., laser fabrication conditions) and the output property (e.g., the observed material structure as labeled by the experts), an optimization procedure is defined and solved.

ML has already revolutionised research fields such as computer vision, speech recognition and natural language processing. The recent success of ML and especially of artificial neural networks stems from their ability to produce super-human performance on tasks where labeled data are abundant \cite{Goodfellow-et-al-2016}. Unfortunately, a large portion of datasets in various scientific and engineering fields have relatively small sizes making the use of data-hungry approaches challenging if not prohibitive due to high generalization errors. As already mentioned, collecting experimental data and to some extent simulated data in laser manufacturing is costly both in time and budget hindering the effort to efficiently and automatically morph the desired structural material properties.
Furthermore, in order to reliably extrapolate the existing knowledge to unknown parameter regimes or novel materials, it is important to develop models that are robust. In principle, simpler ML models are expected to be more robust and they often transfer to unseen parameter regimes. In contrast, complex ML models with no induced constraints not only require more data to be sufficiently trained without being overfitted but also they tend to produce completely off forecasts when applied to an unseen parameter regime.

In this paper, we aim to train and evaluate a series of predictive models with different levels of complexity and expression. Following the Occam’s razor principle, we search for the simplest ML models which do not compromise in terms of predictive performance. The studied ML models are trained and evaluated on three materials (two metals and one semi-conductor) where both experimental and simulated data are available. Performance results indicate that simpler predictive models with a curated and well-educated preprocessing step enjoyed the highest accuracy. This result is a consequence of the amount of training data, which is a crucial factor. In a second stage, the experimental data are further augmented with simulation data extracted from multiscale modeling on the physical processes upon laser-material interaction. This possibility further enabled the study of the effect of sample size on predictive performance, upon retraining the ML models. We observe that the use of a larger sample size, especially, at or close to the transition boundaries, significantly benefits the predictive models’ accuracy. Finally, we propose to quantify the uncertainty regions where the predictive models are not certain about the outcome. Since the studied ML models generate a probability distribution of the structures for each sample, uncertainty quantification is calculated using the information entropy function \cite{Shannon48,Cover2006}. The inverse of the information entropy is used as a measure of certainty.

This paper is organised as follows: in Section \ref{mat:mod:sec}, a basic description of the fundamental physical processes that induce surface modification on solids upon irradiation with fs laser pulses is presented. Section \ref{pred:mod:sec} presents the datasets, the preprocessing steps applied to the input and it briefly introduces and discusses the main characteristics of the predictive models trained and evaluated in this study. The obtained results are demonstrated, analysed and discussed in Section \ref{results:sec} while conclusions are summarised in Section \ref{concl:sec}.

\section{Materials Modeling}
\label{mat:mod:sec}

\subsection{Laser-induced periodic surface structures}
The employment of ultra-short pulsed laser sources for material processing has received considerable attention over the
past decades due to the important technological applications, particularly in industry and medicine \cite{vorobyev2013direct,zorba2008biomimetic,bauerle2013laser,papadopoulou2010silicon,simitzi2015laser,papadopoulos2019biomimetic,stratakis2012nanomaterials}. There is a
plethora of surface structures generated by laser pulses while the so-called laser-induced periodic surface structures
(LIPSS) on solids that have been studied extensively  %\cite{tsibidis2012dynamics,tsibidis2015ripples,huang2009origin,tsibidis2016convection,shimotsuma2003%self,tsibidis2017ripple,bonse2005structure,papadopoulos2018formation,tsibidis2018modelling,birnbaum1%965semiconductor,bonse2016laser,sipe1983laser,tsibidis2020SR} 
are related to those applications. A range 
of LIPSS types have been produced based on the laser parameters %\cite{papadopoulos2019biomimetic,huang2009origin,bonse2016laser,skoulas2017biomimetic,nivas2018surfa%e} 
and the irradiated material. According
to the morphological features of the induced surface structures such as their periodicity and orientation, LIPSS can
be classified in: (a) High Spatial Frequency LIPSS (HSFL), %\cite{bonse2016laser,rudenko2017spontaneous}, 
(b) Low Spatial Frequency LIPSS (LSFL), 
%\cite{tsibidis2012dynamics,bonse2012femtosecond},
(c) Grooves, %\cite{tsibidis2015ripples,nivas2018surface} 
(d) Spikes, 
%\cite{tsibidis2015ripples} 
and (e) complex ones. 
%\cite{skoulas2017biomimetic,nivas2018surface,tsibidis2015ripple}
 The LIPSS fabrication technique as well as the
associated laser driven physical phenomena have been the topic of an extensive investigation. This is due to the fact that
the technique constitutes a precise, single-step and scalable method to fabricate highly ordered, multi-directional and complex surface structures that mimic the unique morphological features of certain species found in nature, an approach
which is usually coined as biomimetics. A thorough knowledge of the fundamental mechanisms that lead to the LIPSS
formation provides the possibility of generating numerous and unique surface biomimetic structures \cite{zorba2008biomimetic,papadopoulos2019biomimetic,skoulas2017biomimetic,kirner2017mimicking,muller2016bio,baron2018biomimetic} for a range of applications including tribology, tissue engineering, advanced optics (for a review on LIPSS and potential applications, see Refs.\cite{bonse2016laser,stratakis2020_review}).

%\cite{vorobyev2013direct,stratakis2011biomimetic}, tribology \cite{bonse2014femtosecond,lu2014biomimetic,wang2016angle}, tissue engineering \cite{simitzi2015laser,stratakis2011biomimetic} and advanced optics \cite{papadopoulos2019biomimetic,jiang2016bioinspired}.

\subsection{Modelling LIPSS}
Various mechanisms have been proposed to account for the development of LIPSS \cite{bonse2016laser,stratakis2020_review}: interference of the incident wave with an induced scattered far-field wave, %\cite{tsibidis2016convection,bonse2005structure,sipe1983laser,rudenko2017spontaneous}
or with a surface plasmon wave (SPW), 
%\cite{tsibidis2012dynamics,tsibidis2014controlled,derrien2013possible,huang2009origin,bonse2009role,barberoglou2013influence}
or due to self-organisation mechanisms. Laser irradiation of solids with ultrashort-pulses involves a series of multiscale processes while the type of structures that are induced is dependent on the laser energy and the energy dose (i.e. number of pulses, $NP$). To describe surface modification for semiconductors and metals, which is the scope of the present study, it is noted that the first pulse leads to the formation of a crater with a rimmed region at the edges for $NP$=1 as a result of possible mass removal (i.e. ablation) and mass displacement \cite{tsibidis2012dynamics}. To evaluate the role of the laser parameters in the surface modification processes, a multiscale modelling approach is required. In principle, modules that account for (i) Laser energy absorption, (ii) Electron excitation, (iii) Heat transfer, (iv) Phase transformation, (v) Resolidification have to be incorporated in the model. These processes occur at different timescales and therefore, to describe, appropriately, the system dynamics, the use of current or revised theoretical models need to be employed. More specifically, the laser energy absorption can be described by approximate solutions \cite{tsibidis2012dynamics} or through a more accurate description of the propagation of the electromagnetic wave of the laser beam \cite{rudenko2016random}.  Electron excitation and dynamics can be modelled through the employment of models based on Boltzmann transport equations \cite{rethfeld2017modelling} or through the use of density functional theory \cite{tsibidis2020modeling,marini2009yambo} while relaxation processes and transfer of the energy of the electron subsystem to the material lattice are described through well-established two temperature models \cite{1974ZhETF..66..776A}. Finally, phase transformation which includes either a transition from solid to the liquid phase and versa or elastic/plastic deformation of a part of the material can be described either through Navier-Stokes \cite{tsibidis2012dynamics} or elastodynamics equations \cite{tsibidis2017ripple,tsibidis2018modelling}, respectively, or from more advanced methods based on Molecular dynamics \cite{ivanov2010nanocrystalline} (for a detailed description of the various modules of the multiscale model see \cite{tsibidis2012dynamics}). Nevertheless, while the aforementioned models are aimed to provide consistent solutions, the picture becomes more complex due to potential limitations of the validity of each particular model. Thus, the use of particular models is especially sensitive to the laser parameter values (i.e. fluence, pulse duration, laser wavelength, polarisation state, material) while coupling of individual modules on each temporal regime, again, is possible to lead to inconsistent and incorrect results.

\begin{figure*}
     \centering
    \subfloat{{\includegraphics[width=0.33\textwidth]{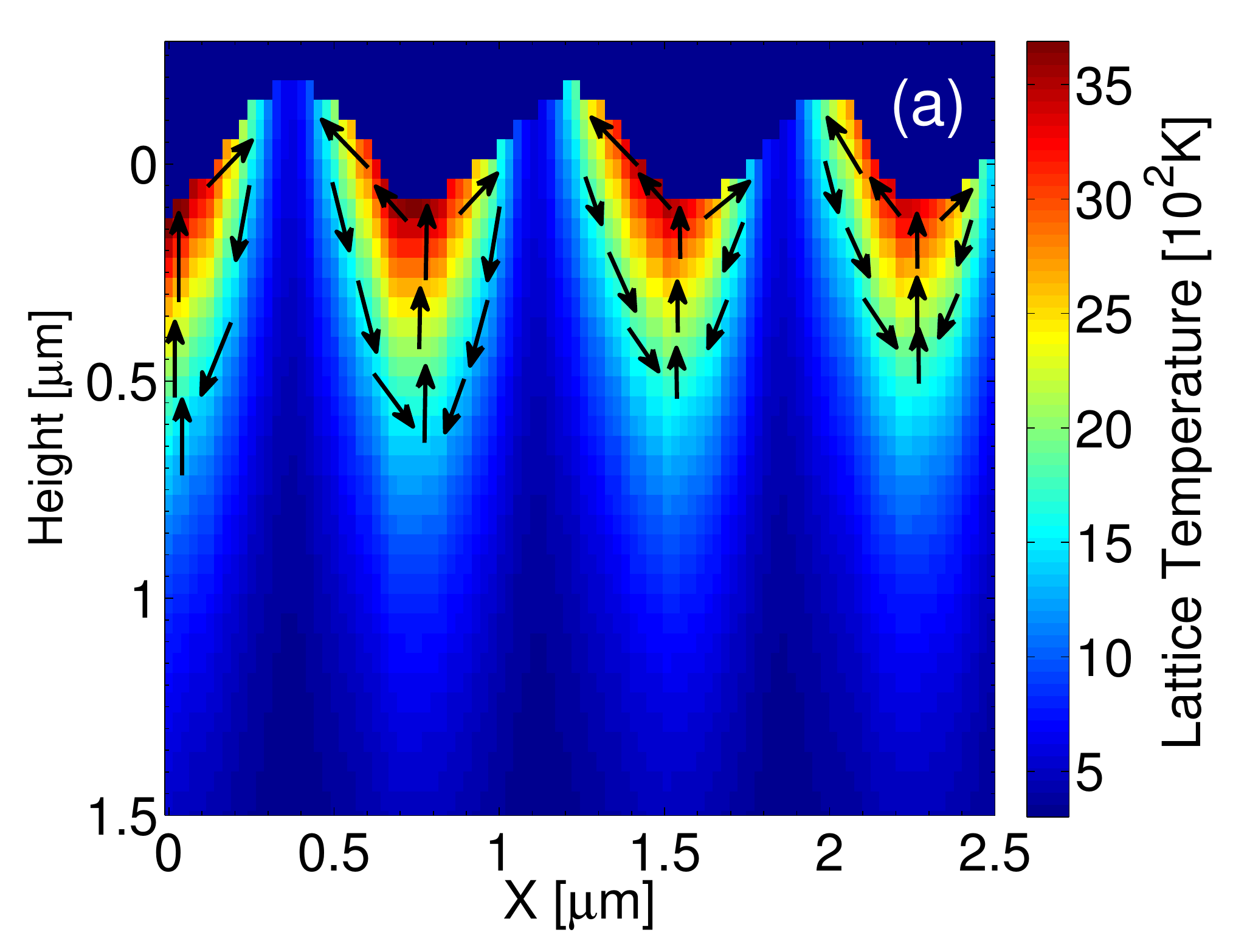} }}%
    %\qquad
    \subfloat{{\includegraphics[width=0.33\textwidth]{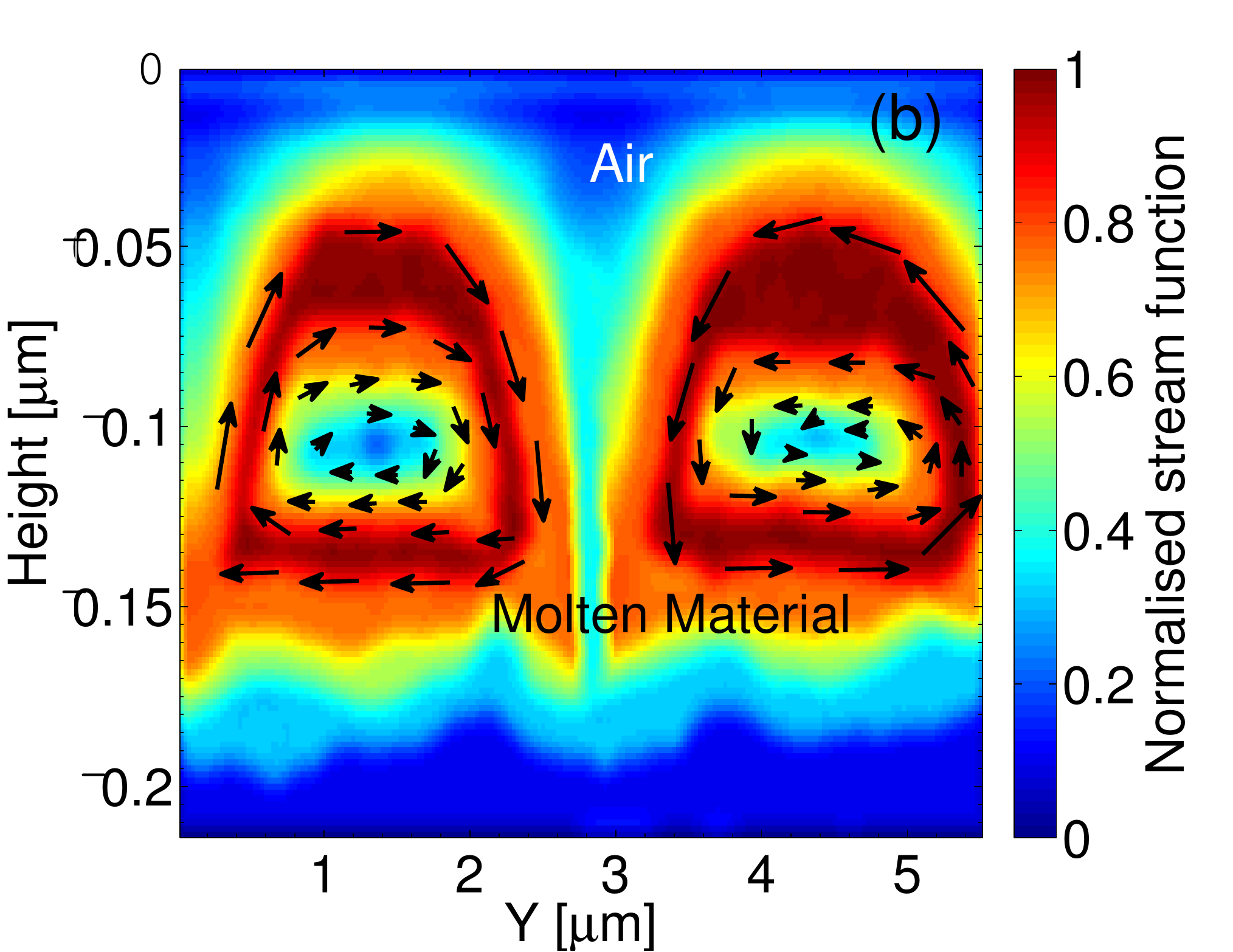} }}%
    %\qquad
    \subfloat{{\includegraphics[width=0.33\textwidth]{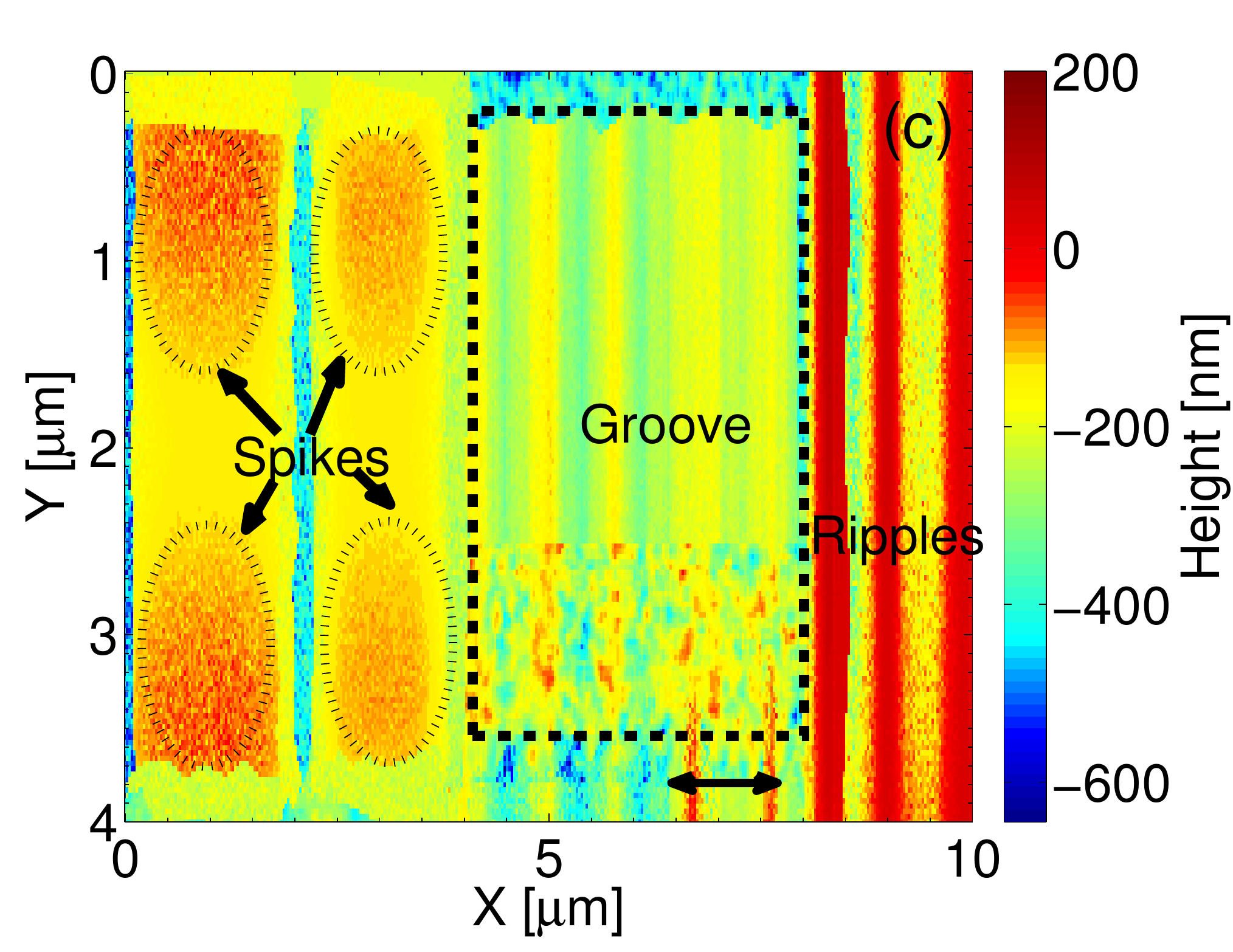} }}%
    \caption{(a) Fluid Transport, (b) Convection roll formation, (c) patterned surface. Double-ended arrow in (c) indicates the laser beam polarisation. [Reproduced with permission from Tsibidis et al. \cite{tsibidis2015ripples}. Copyright (2015) by the American Physical Society].}%
    \label{fluid:conv:patterned:fig}%
\end{figure*}

While the multiscale model presented above is, in principle, capable to provide the mechanism for surface patterning, appropriate modifications are required to describe accurately morphological changes at increasing $NP$ (or fluence). More specifically, irradiation of a non-flat profile as a result of irradiation with $NP$=1, leads to an interference of the incident beam with a scattered wave (it could involve the excitation of Surface Plasmon wave, SP) resulting, in turn, into a spatially modulated energy deposition on the surface of the material \cite{tsibidis2012dynamics}. The spatially modulated form of the energy distribution yields a periodic shape (with periodicity equal to the SP wavelength \cite{tsibidis2012dynamics,derrien2013possible,bonse2009role}) that is projected firstly, onto the excited electron dynamics, electron temperature and lattice temperature and finally to the longer timescale effects related to melting and resolidification of the irradiated material
\cite{bonse2016laser,stratakis2020_review}.
%\cite{tsibidis2012dynamics,tsibidis2015ripples,tsibidis2016convection,tsibidis2015ripple,barberoglou2013influence,tsibidis2020modeling,margiolakis2018ultrafast,gakovic2017partial}.
As a result, a surface profile covered with LIPSS of periodicity of the size of the wavelength will be created . These structures are orientated perpendicularly to the polarisation vector of the laser beam and are termed as LSFL (for the sake of simplicity, in this report, they will be called $ripples$ \cite{tsibidis2012dynamics}). As $NP$ increases or at higher energies (i.e. fluences, $F$), the profile becomes deeper and the height of the ripple crest increases; various theoretical models have been developed to account for the experimentally observed decrease of the periodicity of LIPSS \cite{huang2009origin,tsibidis2017ripple}. A saturation point is reached when SP excitation ceases after a large number of $NP$; afterwords, melting of the material leads to a fluid transport along the well of the rippled zone leading to another type of structures, the $grooves$ \cite{tsibidis2015ripples}. These structures are orientated parallel to the orientation of the laser beam while their periodicity is larger than the laser wavelength. Further irradiation with larger $NP$ is expected to lead to pointed structures, that are termed $spikes$ \cite{tsibidis2015ripples}. Theoretical results based on multiscale modelling show the fluid transport that leads to the formation of ripples (Fig. \ref{fluid:conv:patterned:fig}(a)) \cite{tsibidis2012dynamics,tsibidis2015ripples} while convention roll development and movement  account for the formation of grooves (Fig. \ref{fluid:conv:patterned:fig}(b)) \cite{tsibidis2015ripples,tsibidis2018modelling}. Finally, spike formation is also explained through the employment of hydrodynamical models (Fig. \ref{fluid:conv:patterned:fig}(c)) \cite{tsibidis2015ripples}.

\subsection{Experimental setup}
\begin{figure*}
\includegraphics[width=\textwidth]{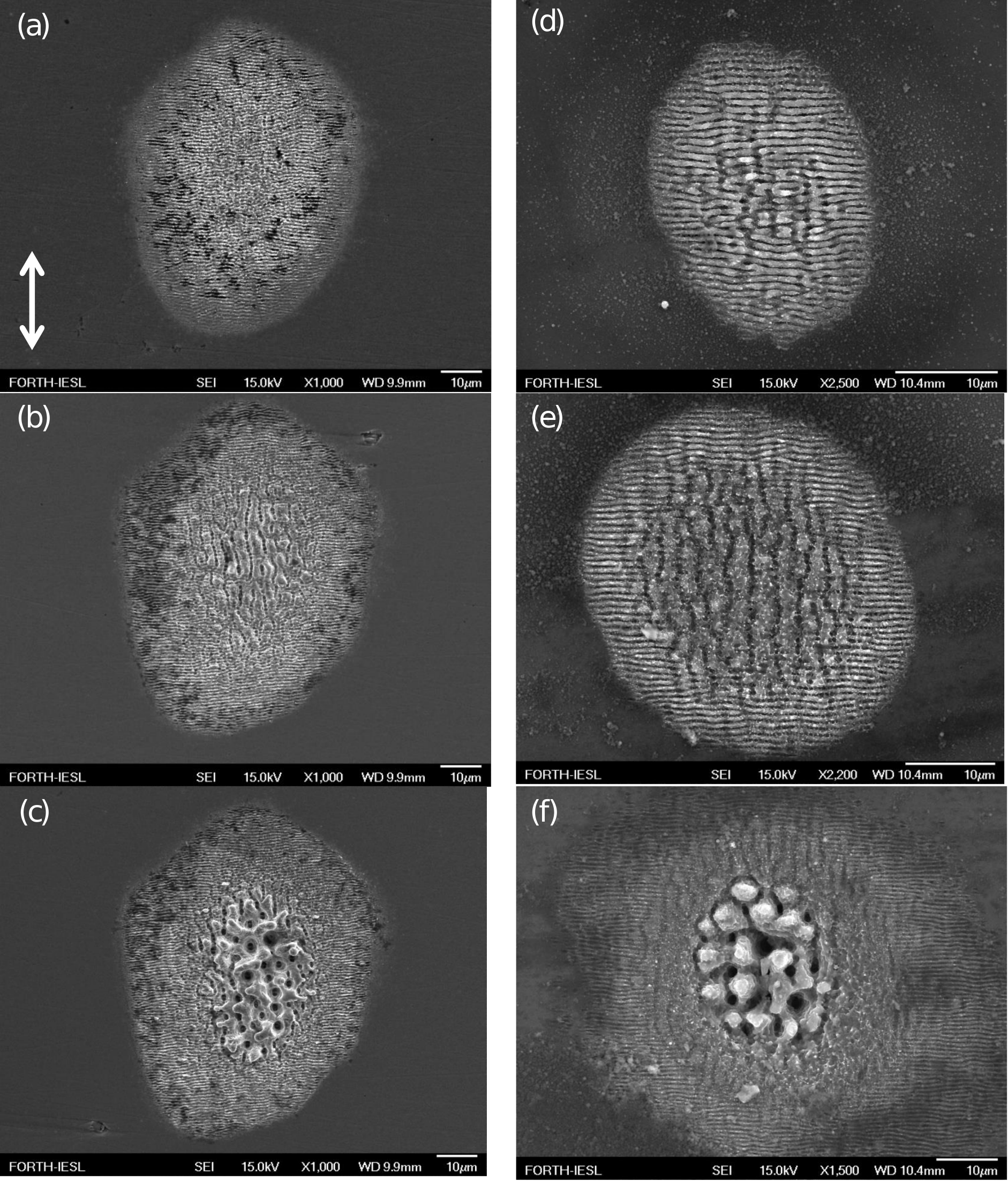}
    \caption{SEM images: (a) Ripples, (b) Grooves, (c) Spikes in the central region for Steel,(d) Ripples, (e) Grooves, (f) Spikes in the central region for Silicon. Double-headed arrow indicates laser beam polarisation orientation}%
    \label{sem:images:fig}%
\end{figure*}

Laser irradiation is performed with the use of an Yb:KGW Pharos – SP laser system from Light Conversion that emits linearly polarized IR pulses at 1026 nm central wavelength at 1 kHz repetition rate of 170 fs pulse width. The samples used in the experimements were specimen of stainless steel ( 1.7131), alpha-beta Titanium alloy (Ti6Al4V) and crystalline Silicon (Si) ($p$-doped). The laser beam was guided using silver mirrors and focused on the sample surface with an focal length $f$=200 mm plano convex lens. The
spot size was characterized with a CCD camera close to the focal plane and was estimated around $\sim 60 \mu m$ and consistent with a Gaussian intensity profile. Irradiation was performed within the Rayleigh range of the focal position and the number of pulses receptive to the sample for static irradiation were controlled with an external mechanical shutter.
The laser power was modulated from the laser amplifier settings and all irradiation experiments were performed at normal incidence. Fig.\ref{sem:images:fig} shows Scanning Electron Microscopy (SEM) images for the surface patterns obtained for stainless steel (left column) and Silicon (right column) in which formation of ripples, grooves or spikes in the central region is visible. To emphasise on the role of $NP$ and $F$ in the formation of various types of structures, the impact of both the fluence and energy dose is illustrated in Fig.\ref{sem:images:fig}(a) ($NP$=2, $F$=1.5 J/cm$^{2}$), Fig.\ref{sem:images:fig}(b) ($NP$=40, $F$=1.5 J/cm$^{2}$), Fig.\ref{sem:images:fig}(c) ($NP$=80, $F$=1.5 J/cm$^{2}$), Fig.\ref{sem:images:fig}(d) ($NP$=80, $F$=0.4 J/cm$^{2}$), Fig.\ref{sem:images:fig}(e) ($NP$=80, $F$=0.7 J/cm$^{2}$), and Fig.\ref{sem:images:fig}(f) ($NP$=80, $F$=1.5 J/cm$^{2}$),

\subsection{Materials modelling simulation approach}
\begin{figure*}
     \centering
   \includegraphics[width=\textwidth]{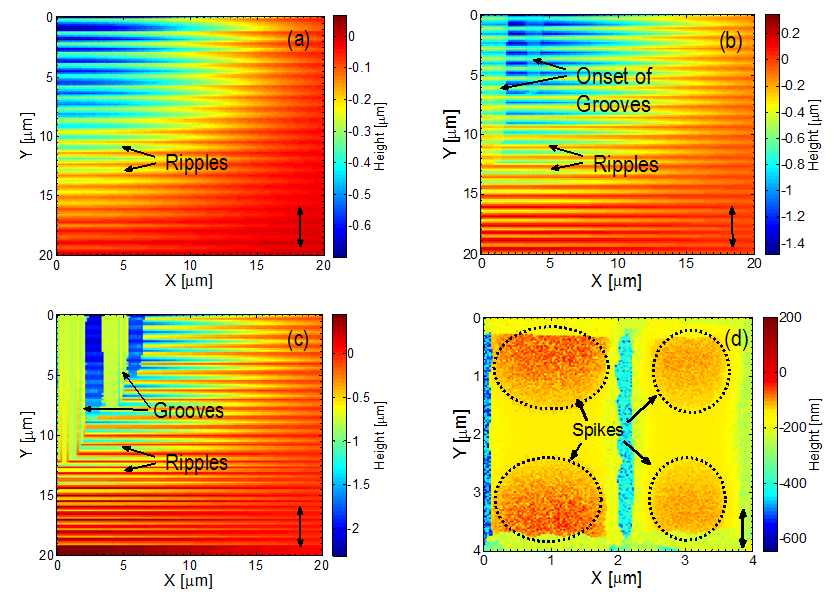}
    \caption{Top view of a quadrant of simulated surface pattern for Silicon (a) Ripples ($NP$=10), (b) Hybrid region with ripples and grooves and Ripples ( $NP$=60), (c) Grooves and Ripples ( $NP$=100), and (d) Spikes ( $NP$=400). $F$=1.4 J/cm$^{2}$. Double-headed arrow indicates laser beam polarisation orientation}
    \label{simulated:regions:fig}%
\end{figure*}

A common approach followed to solve the set of equations constituting the multiscale model (describing laser energy absorption, electron excitation, heat transfer and relaxation processes, hydrodynamics, resolidification and elastoplasticity) \cite{tsibidis2015ripples, tsibidis2012dynamics} is the employment of a staggered grid finite difference method which is found to be effective in suppressing numerical oscillations \cite{tsibidis2015ripples, tsibidis2012dynamics,tsibidis2015ripple}. Unlike the conventional finite difference method, temperatures (heat transfer equations), electron densities, and pressure are computed at the centre of each element while time derivatives of the displacements and first-order spatial derivative terms are evaluated at locations midway between consecutive grid points. For time-dependent flows, a common technique to solve the Navier-Stokes equations (for fluid transport) is the projection method and the velocity and pressure fields are calculated on a staggered grid using fully implicit formulations \cite{tsibidis2015ripples, tsibidis2012dynamics,tsibidis2015ripple} . On the other hand, the horizontal and vertical velocities are defined in the centres of the horizontal and vertical cells faces, respectively. A multiple pulse irradiation scheme is required to derive the  surface relief \cite{tsibidis2015ripples, tsibidis2012dynamics,tsibidis2015ripple} and therefore after each $NP$, the induced profile is used to compute the energy absorption and dynamics when the next pulse irradiates the material. To simulate the multiscale process, a numerical scheme based on the use of a finite difference method is followed while the discretisation of time and space has been chosen to satisfy the Neumann stability criterion. It is assumed that on the boundaries, von Neumann boundary conditions are satisfied and heat losses at the front and back surfaces of the material are negligible. The initial conditions are both the electron and lattice temperatures are at room temperature. To simulate mass removal, it is assumed that it occurs if the material is heated above a critical temperature. The boiling temperature of the material is selected as the critical temperature. Simulations were conducted for all four materials. The simulated surface patterns for $NP$=10 (ripples) and $NP$=100 (grooves) for Silicon are illustrated in Fig. \ref{simulated:regions:fig}(a) and Fig. \ref{simulated:regions:fig}(c), respectively, for $F$ =1.4 J/cm$^{2}$ while at lower $NP$ (i.e. $NP$=60) the distinction between groove formation and ripple suppression in the central region is not clear (Fig. \ref{simulated:regions:fig}(b)). The latter structures resemble the early stage of groove formation. Similar results have been produced for the three other materials. Finally, Fig. \ref{simulated:regions:fig}(d) illustrates spike formation for $NP$=400.

\section{Predictive Modeling}
\label{pred:mod:sec}
In this section, we describe the annotated datasets and their statistical properties as well as the proposed preprocessing steps. We also present several standard ML models along with their advantages and disadvantages.
\subsection{Material Datasets}

\begin{figure*}
\includegraphics[width=\textwidth]{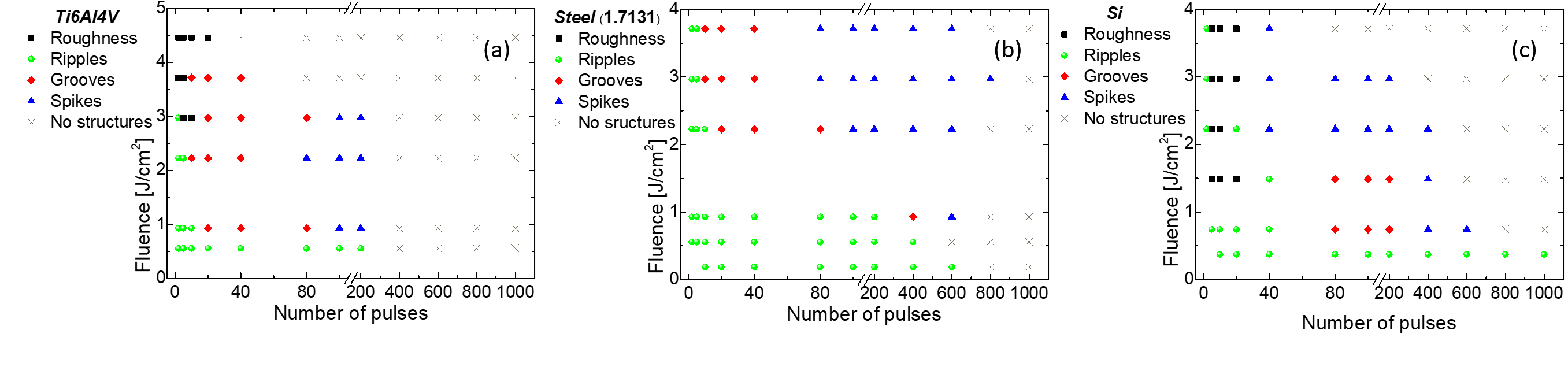}
     \caption{Experimental results: Morphological maps for (a) Ti6Al4V, (b) Stainless Steel (1.7131), and (c) Si. }%
    \label{exp:data:fig}%
\end{figure*}

\begin{figure*}
\includegraphics[width=\textwidth]{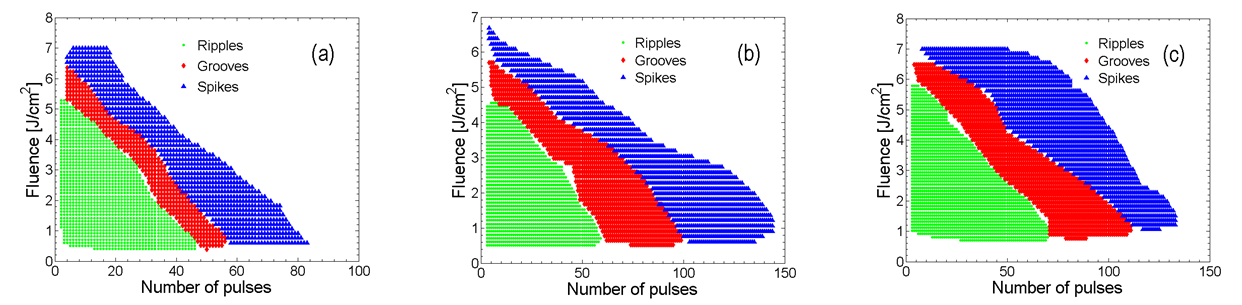}
    \caption{Simulation results: Morphological maps for (a) Ti6Al4V, (b) Stainless Steel (1.7131), and (c) Si.  }%
    \label{sim:data:fig}%
\end{figure*}

A library of structure types and samples is acquired as a function of $F$ and $NP$. Figs. \ref{exp:data:fig} and \ref{sim:data:fig} show the distributions of the annotated (i.e., labeled with the structure type) experimental and simulated data, respectively, for three materials. The dependence of the structure types on the laser conditions is clear in all cases. Moreover, a comparison of the structure maps produced with simulations and experiments show that more types of classification are observed during experiments (i.e. termed as 'roughness' or 'no structures') which are not predicted in simulations. Furthermore, the maps indicate that the onset of one type of structure does not occur for the same combination of the fluence and the energy dose. These effects can be attributed to the fact that the physical models that are used to describe the underlying processes aim to approximately account for the physical mechanisms that take place; the validity of those models are, thus, characterised by limitations and revised theoretical frameworks might be required. On the other hand, laser patterning conditions in an experimental protocol cannot be easily specified to agree with the values selected in the theoretical models. This ambiguity is, thus, expected to influence the energy deposition and the laser absorbed energy which is critical for the production of the structure types. 

It is also evident that a significant number of experiments (or simulations) is required for the accurate determination of the structures. Tables \ref{experimental:table} \& \ref{simulated:table} report the number of samples per structural category per material for the experimental and simulation datasets, respectively. The small number of experimental samples cannot be ignored during the selection and training of predictive models. Finally, it is worth-noting that special attention is required around the boundary regions where the transition from one structure to another occurs (i.e. these hybrid states that are described in the previous sections are illustrated as white zones between regions). Hence, an improved precision requires a larger number of time consuming simulations and/or experiments. Therefore, the development of an alternative, systematic, less computational expensive and reliable predictive tool is needed to determine the structure regions.  

\begin{table}[h]
\caption{Number of samples for each structure and each material for the experimental data.}
\begin{ruledtabular}
\begin{tabular}{cccccc}
& Total & Ripples & Grooves & Spikes & Other\\
\hline
Si & 68 & 19 & 6 & 13 & 30\\
Ti6Al4V & 72 & 14 & 12 & 7 & 39 \\
Steel (1.7131) & 70 & 32 & 10 & 16 & 12\\
\end{tabular}
\end{ruledtabular}
\label{experimental:table}
\end{table}

\begin{table}[h]
\caption{Number of samples for each structure and each material for the simulated data.}
\begin{ruledtabular}
\begin{tabular}{ccccc}
& Total & Ripples & Grooves & Spikes\\
\hline
Si & 4886 & 1307 & 1951 & 1628\\
Ti6Al4V & 2502 & 432 & 1244 & 826 \\
Steel (1.7131) & 3964 & 1141 & 1482 & 1341 \\
\end{tabular}
\end{ruledtabular}
\label{simulated:table}
\end{table}

\subsection{Feature Construction}
Despite the development of several general purpose ML models during the last decades, not all models are appropriate for all data collections since there are trade-offs between dataset size and complexity as well as between accuracy and robustness or transferability to new data. Especially, in applications where data are relatively scarce and expensive to generate either less complex models have to be selected or experimental data must be supplemented with simulated data. Another important yet usually hidden aspect of ML success is the proper preparation and preprocessing of the data. Even though ML models are built to work for any type of data distributions, the results are often significantly improved by simple transformation and/or normalization of the data. For instance, various control parameters in many physical/chemical/material systems take values that range in several orders of magnitude. The utilisation of nonlinear transformations such as the power transformation or the application of the logarithmic function regularises the statistical properties of the parameters towards a Gaussian distribution assisting the performance of predictive models. In this paper, we apply the logarithmic transformation to the number of pulses, $NP$, while we keep unchanged the fluence parameter, $F$.

Remembering that a predictive model is a complex mapping between the input and the output, this mapping describes the nonlinear relationships and interactions between the input variables and the output. On the other hand, it is often beneficial to reduce the complexity of the mapping, hence, of the model whenever this is possible by applying feature construction. A standard approach to construct new features from existing ones is by taking the product or the ratio between two or more features. In this work, we take the product between two and three features and construct the quadratic as well as the cubic terms. Given that the initial feature vector has two elements (logarithm of $NP$ and $F$), there are three quadratic and four cubic terms resulting in three datasets per material with two, five and nine features, respectively.

\subsection{Machine Learning Models}
We will describe, briefly, the features of some representative predictive models and their characteristic properties. Typically, a predictive model has a trainable set of coefficients (or parameters), $\theta$, and it approximates the posterior probability distribution $P(c|x)$ of each label, $c$, given the sample, $x$ \cite{murphy2013machine}. The label corresponds to the material structure while the sample corresponds to the fabrication conditions (or configuration). After training, the structure prediction for a new configuration, $x'$, is obtained from
\[
c' = \arg\max_c P_{\hat{\theta}}(c|x')
\]
where $\hat{\theta}$ is the learned coefficient vector. We start the demonstration with the simpler models and then proceed with more advanced ones aiming to clarify which models are appropriate based on their ability to learn from small sample sizes, their complexity and interpretability.
\newline

% \subsubsection{$k$ Nearest Neighbors ($k$-NN)}

\noindent
{\bf $k$ Nearest Neighbors ($k$-NN).}
$k$-NN is an instance-based classification method where a new instance (i.e., sample) is compared with existing instances already available in the training set \cite{DudaHartStork01,bishop2007}. $k$-NN decides the label of new samples from the labels' occurrence frequency of the $k$ closest neighboring samples. There is no training involved in $k$-NN and the user has to specify the number of neighbors, $k$, and define the distance (or similarity metric) between the samples. The distance must take into account the type of the features (discrete, ordinal, continuous, etc.). The performance of $k$-NN is sensitive to the chosen distance and it might be heavily deteriorated by an poorly-behaved distance. For instance, features with high variance could dominate the distance value resulting in bad performance. Thus, data preprocessing techniques such as feature normalization is typically required. $k$-NN serves as a standard baseline model when the number of samples is low, however, it does not scale well with the feature dimension or for large datasets where the inference is proportional to the sample size making $k$-NN very slow.
\newline

%K-NN is an Instance-based learning meaning it compares new problem seen in training, which are stored in memory. The classification of the data is determined by a set of previously classified points. More specifically the k nearest points in the feature space decide the class through a plurality vote.

% \subsubsection{Gaussian Naive Bayes (GNB)}

\noindent
{\bf Gaussian Naive Bayes (GNB).}
Naive Bayes method is a probabilistic classifier which assumes that each feature is independent from any other \cite{DudaHartStork01,bishop2007}. Under this strong independence assumption, the posterior probability is straightforwardly calculated using the Bayes theorem. GNB additionally assumes that the conditional univariate random variables follow the normal distribution parametrized by the mean and the variance. GNB's parameter estimation, which is fast, is typically performed using the maximum likelihood method. The simplicity of GNB model often results in inferior results however it has better generalization performance when the independence assumption stands true. GNB is also utilized as a baseline model that offers a point of reference for more sophisticated ML approaches.
\newline

%It takes advantage the Naive Bayes theorem which is based on conditional probability where we calculate the probability that something will happen given that some else has already happen. In Gaussian NB model we assume that the features have continues values and that are following a Gaussian distribution.

% \subsubsection{Logistic Regression Model (LRM)}
\noindent
{\bf Logistic Regression Model (LRM).}
Logistic model assigns the probability of a particular label in a binary classification problem \cite{hastie01statisticallearning}. It belongs to the family of generalized linear models where the output is a non-linear function of a linear combination of the inputs. The non-linear function known as the logit function (from logistic unit) is statistics outputs the probability of the input to belong to one of the two classes. The regression coefficients are estimated using maximum likelihood estimation. There are no hyper-parameters for tuning making LRM easier to train. The LRM formulation can be straightforwardly generalized to multi-labeled classification problems. 

Moreover, LRM is interpretable in the sense that the regression coefficient contain information about the importance of the respective feature. This is an important advantage over most ML models which are treated as black boxes since it can lead to knowledge discovery. Enhanced interpretability is achieved when maximum likelihood estimation is supplemented with a regularization term that favors parsimonious models \cite{tibshirani96regression, hastieElasticNet}. Sparse models can reveal the dominant and relevant features that correlate with the outcome. 
\newline

%Logistic Regression can be used in 3 different types of classification. First in simple binary problems (e.g.yes/no question). Taking it 1 step further we have multiclass problems. And lastly we have ordinal logistic regression where the classes have some kind of ordering (e.g. ratings from 1 to 5). Logistic Regression uses a Sigmoid function which takes any real value between zero and one. it is defined as $\sigma(x)=1/(1+exp(-\beta^T x))$. Then a decision boundary is set to differentiate probabilities into positive class and negative class. 

% \subsubsection{Support Vector Machines (SVM)}
\noindent
{\bf Support Vector Machines (SVM).}
An SVM classifier aims to separate the data points into two classes by constructing a maximum-distance hyperplane \cite{cortes1995support,Scholkopf2002,cristianini2000introduction}. The data points that participate in the construction of the hyperplane are called support vectors. Utilizing the kernel method, data are transformed in high or even infinite dimensions where the optimal boundary is searched for. The optimization problem is solved using quadratic programming \cite{Scholkopf2002}. The user has to specify which kernel is most appropriate for the task at hand as well as the soft margin parameter that controls how much misclassification error is allowed. SVM training scales as a function of sample size with an order somewhere between quadratic and cubic making them computationally expensive for large datasets. Nevertheless, SVMs frequently produce state-of-the-art results for low sample size datasets \cite{ICML-2006-CaruanaN}.
\newline

%SVM is a supervised machine learning algorithm which can be used for classification. It uses the kernel method to transform the data and then it finds an optimal boundary. More specifically it defines a hyperplane which sections the data in 2 subsets each one representing a class. We define the points closer to that hyperplane as the support vectors.

% \subsubsection{Gradient Boosting Classifier (GBC)}
\noindent
{\bf Gradient Boosting Classifier (GBC).}
Gradient Boosting is an ML technique that iteratively and gradually learns the classification task from many weak classifiers \cite{freund95decisiontheoretic,friedman2000greedy}. Typically, the weak classifier is a classification and regression tree \cite{BreFriOlsSto84a} which represents a decision making process. GBC builds a bag of trees one at a time, where each new tree aims to eliminate errors made by the already trained trees and at the end it produces an ensemble model determined by the weak classifiers. Gradient boosting is an optimization problem in the sense that a loss function is calculated and optimised in each step \cite{ChenG16}. The typical loss corresponds to a function of the residual errors (the difference between actual value and predicted value) which is iteratively minimized. GBC is sensitive to noise in the data and relatively harder to tune, however, it has generated excellent results in a wide spectrum of ML tasks \cite{ICML-2006-CaruanaN} and it is one of the most popular ML techniques. 
\newline

%In decision analysis, a decision tree can be used to represent decision making. The gradient boosting classifier produces a prediction model using an ensemble of weak prediction models, typically decision trees. We can see gradient boosting as an optimization problem meaning that in each step we calculate a loss function and try to optimize it. So we iteratively we add another weak learner and compute the loss. The loss represents the error residuals (the difference between actual value and predicted value) and using this loss value the predictions are updated to minimize the residuals.

% \subsubsection{Additional ML Models}
\noindent
{\bf Additional ML Models.}
The resurrection of artificial neural networks in the recent years has been attributed to their excellent performance both as feature constructors and as powerful classifiers \cite{Goodfellow-et-al-2016}. The potential of neural nets in laser manufacturing is unquestionable especially when the target is to classify material properties from raw data such as microscope images. However, the learning task we study here does not require models with such large learning capacity as neural nets. Nevertheless, we present a series of results with neural nets in Appendix showing that they do learn the studied classification task given that we provide sufficiently many samples for training.

Finally, another popular ensemble model forth mentioning is random forests \cite{breiman2001random} whose building blocks are decision trees as with GBC. However, there are significant differences between GBC and random forests in terms of how the trees are built as well as on how the results of each tree are combined. Despite producing state-of-the-art results in several classification tasks over the years, random forests often perform worse than GBC when the hyper-parameters of GBC are carefully tuned \cite{ICML-2006-CaruanaN}.

\section{Results}
\label{results:sec}
The performance assessment of ML models on their ability to correctly predict the material structure given a laser configuration is presented in this Section. We will show the importance of feature construction in both experimental and simulated data as well as the fact that measuring and annotating sufficient amounts of data plays a crucial role for excellent predictive performance.

%In this section we test to see if there is an improvement in our results depending on the features used. Indeed we can see in figure * that we get a lot better results in both experimental and simulated data, using cubic features which might indicate that there is an underlying correlation in some cubic or quadratics feature that needs to be further investigated. It is also worth mentioning that some of the models do not get affected by using quadratic or cubic features. 

\subsection{Performance metrics and evaluation protocol}
We use as a metric of performance the Area Under the ROC\footnote{ROC: Receiving Operating Characteristics from detection theory \cite{Egan75}.} Curve (AUC) which is insensitive to the number of instances per category. Since the classification tasks are multiclass and not binary classification tasks, we choose to convert them into “One vs Rest” binary classification problems where the first class corresponds to data from one label while we merge the remaining data to obtain the second class. Then, we compute the true positive and false positive rates for each class. In order to construct the overall ROC curve for a material, we compute the average for all classes. 

We follow the standard cross validation (CV) protocol and split the data into training and testing datasets. CV is necessary in order to obtain unbiased results. We apply $k$-fold CV where each dataset is split into $k$ subsets and $k-1$ of them are used for training and the remaining for performance assessment. The procedure is repeated $k$ times --one time for each subset-- and the average performance is calculated and reported. Due to the fact that the experimental data are limited, we set $k=n$ where $n$ is the sample size of the dataset, a technique known as Leave One Out CV (LOOCV). We apply 6-fold CV for the simulated datasets. Finally, we performed limited hyper-parameter tuning and the optimal hyper-parameter values are shown in Table \ref{hyperparam:table}.

\begin{table}%[h]
\caption{Optimal hyperparameter (HP) values for each predictive model and both experimental (exp) and simulated (sim) datasets.}
\begin{ruledtabular}
\begin{tabular}{cccc}
&HP name& value (exp)&  value (sim) \\
\hline
    \multirow{2}{*}{SVC}&kernel&linear&linear\\
    % &gamma&10&default\\
    &C&30&default\\
\hline

% \multirow{}{}{LR} &solver&liblinear&lbfgs\\
\multirow{2}{*}{LR} &norm& $l_2$ (default) & $l_2$ (default) \\
&C&30&default\\
\hline
GB&learning rate&0.2&0.1\\
\hline
GNB&-&- &- \\
\hline
KNN&$k$&5& 20 \\
\end{tabular}
\end{ruledtabular}
\label{hyperparam:table}
\end{table}

\subsection{Predictive performance of constructed features}
\begin{figure*}
    \centering
    \subfloat[Results on experimental data.]{{\includegraphics[width=0.4\textwidth]{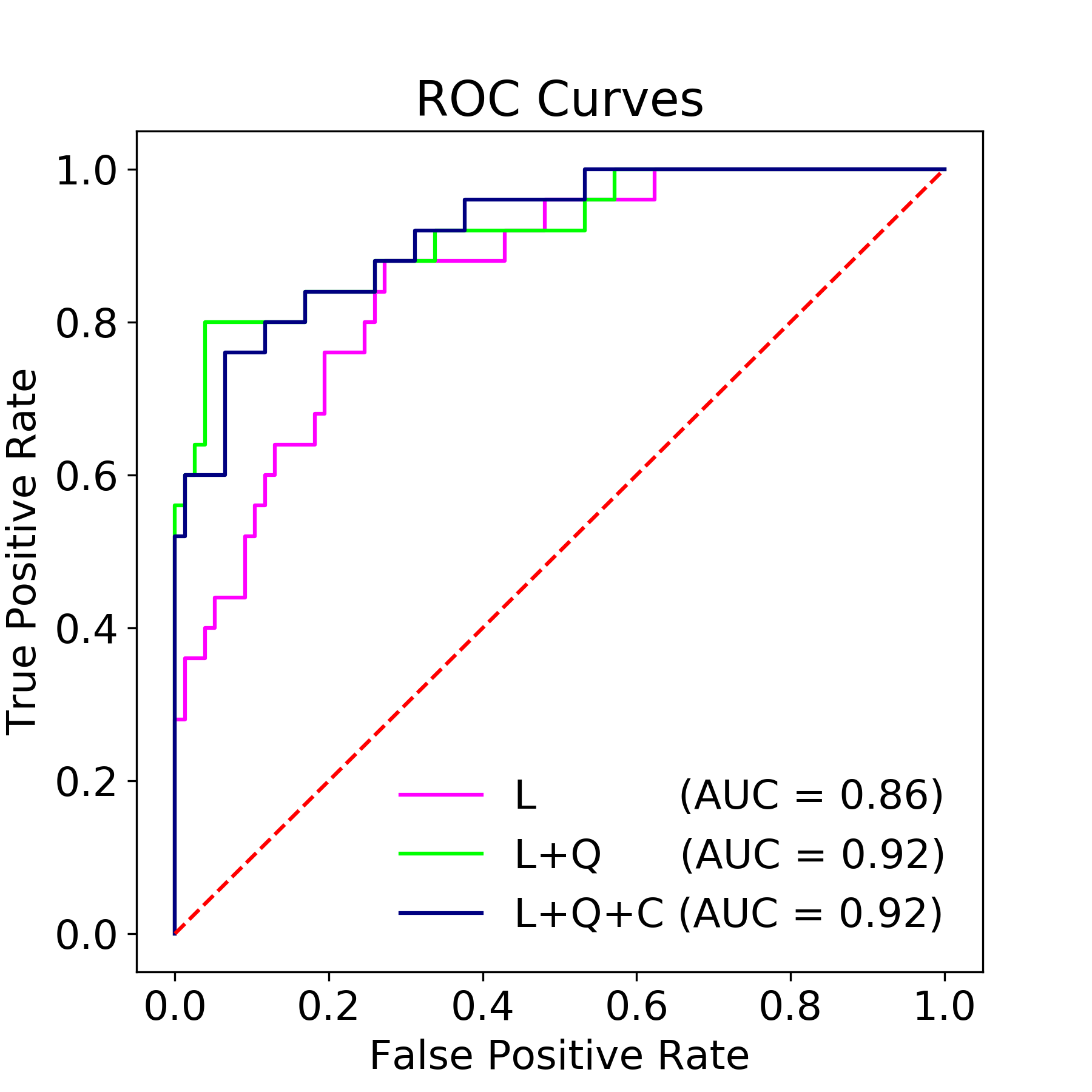} }}%
    \qquad
    \subfloat[Results on simulated data.]{{\includegraphics[width=0.4\textwidth]{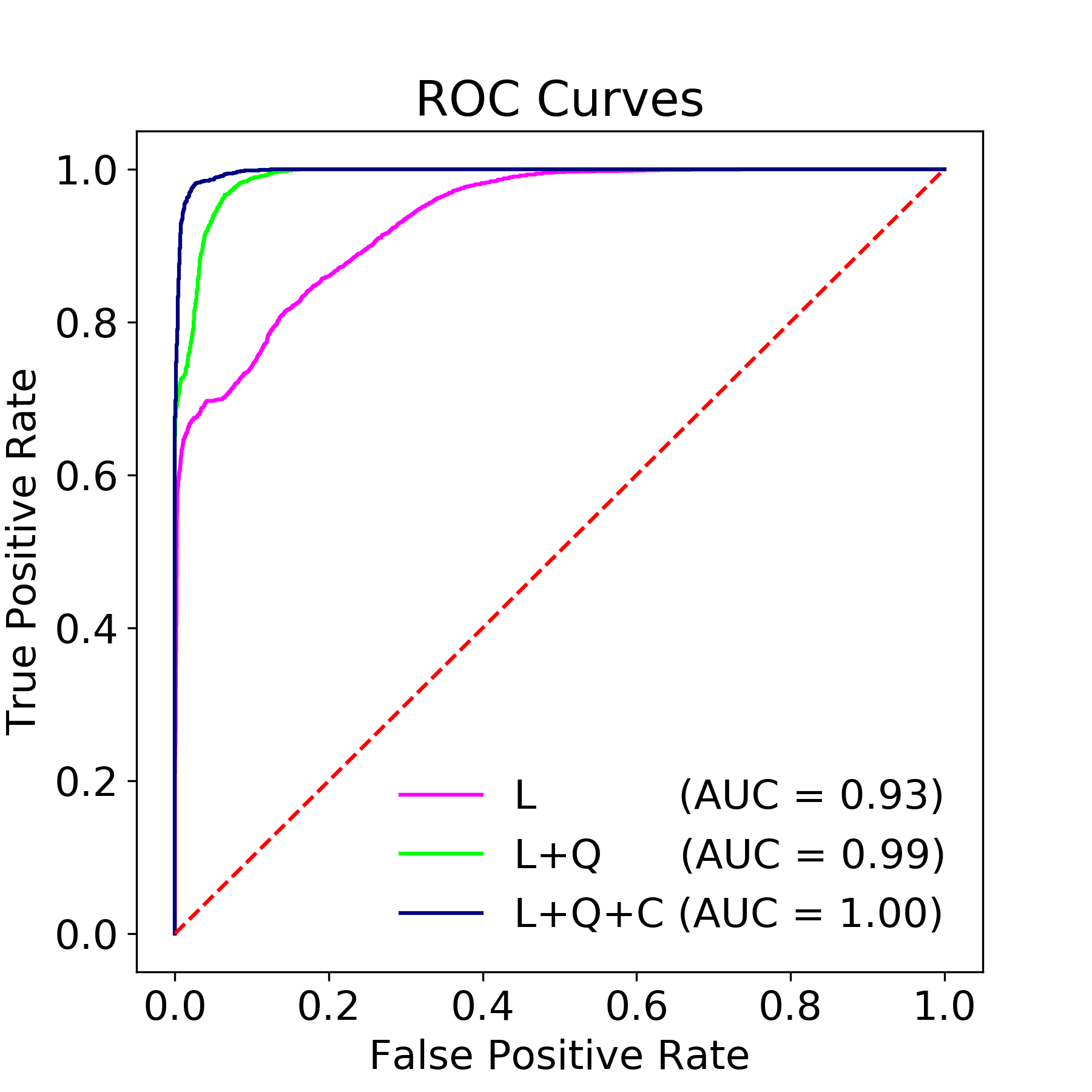} }}%
    \caption{Performance assessment of constructed features for Silicon (Si). The use of nonlinear feature interactions significantly assist the higher accuracy of the LR model.}%
    \label{various:feat:fig}%
\end{figure*}

We first assess the performance of the constructed features in terms of average AUC for the LR model. The use of logistic regression is preferred over nonlinear models such as SVMs because we want to avoid imperil with model biases the predictive power of the additional features (see also Supplementary Materials ). Fig. \ref{various:feat:fig} shows the ROC curves for linear (magenta lines), linear+quadratic (green lines) and linear+quadratic+cubic (blue lines) features for Silicon. The dotted orange line corresponds to the random classifier. Both experimental (left panel) and simulated (right panel) datasets are considered. In both cases, the constructed features assisted the predictive model to increase its accuracy. The average AUC value in experimental datasets is increased by 7\% when the quadratic features are added as well as when both quadratic and cubic features are added as input. On simulated data, LR model achieved the highest average AUC value which is an excellent result showing that the use of nonlinear features along with larger datasets results in better predictive performance. We also observe qualitatively similar results for the other materials of this study.

\subsection{Performance of predictive models on experimental data}
\begin{figure*}
    \centering
    \subfloat[Ti6Al4V]{{\includegraphics[width=0.33\textwidth]{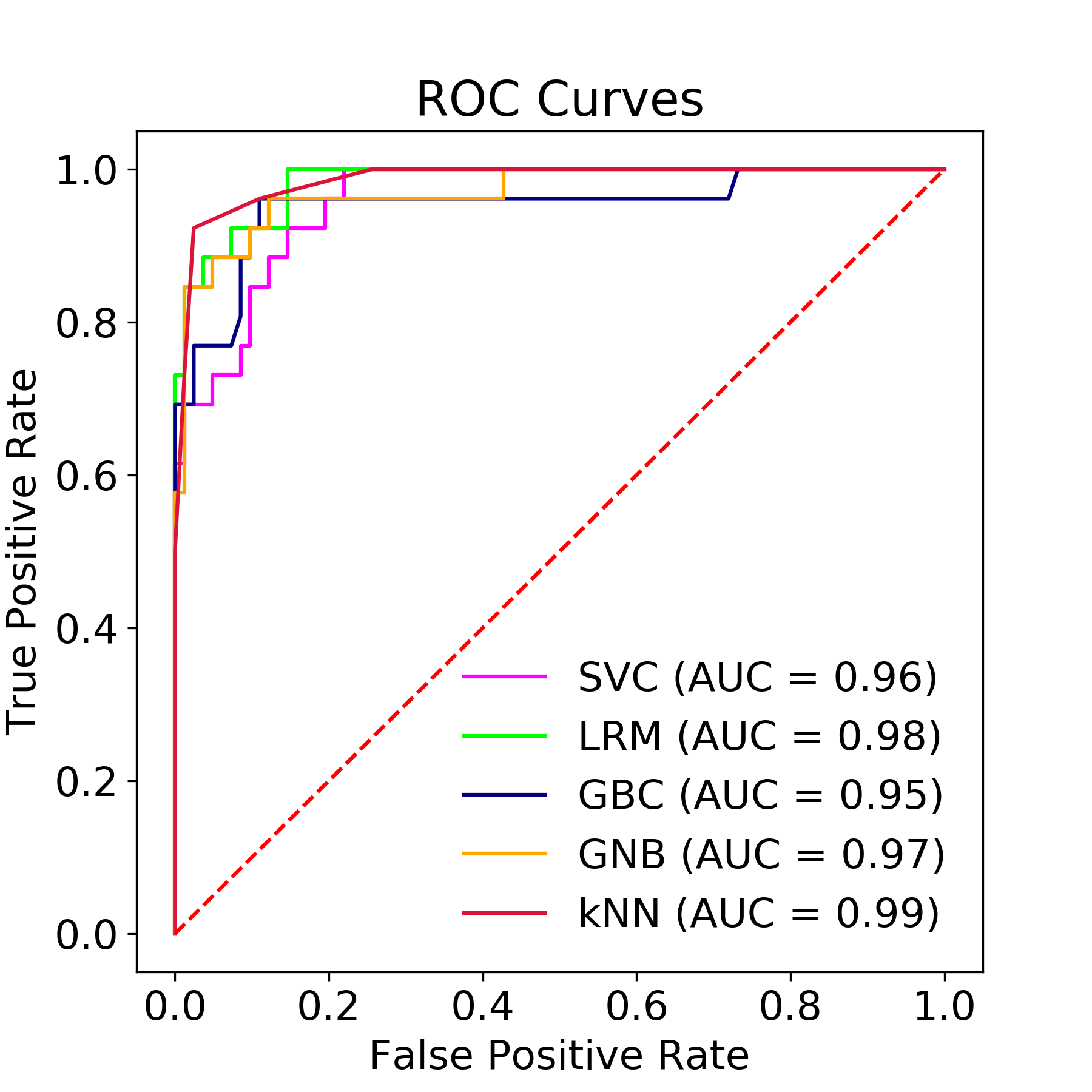} }}%    
    \subfloat[Steel (1.7131)]{{\includegraphics[width=0.33\textwidth]{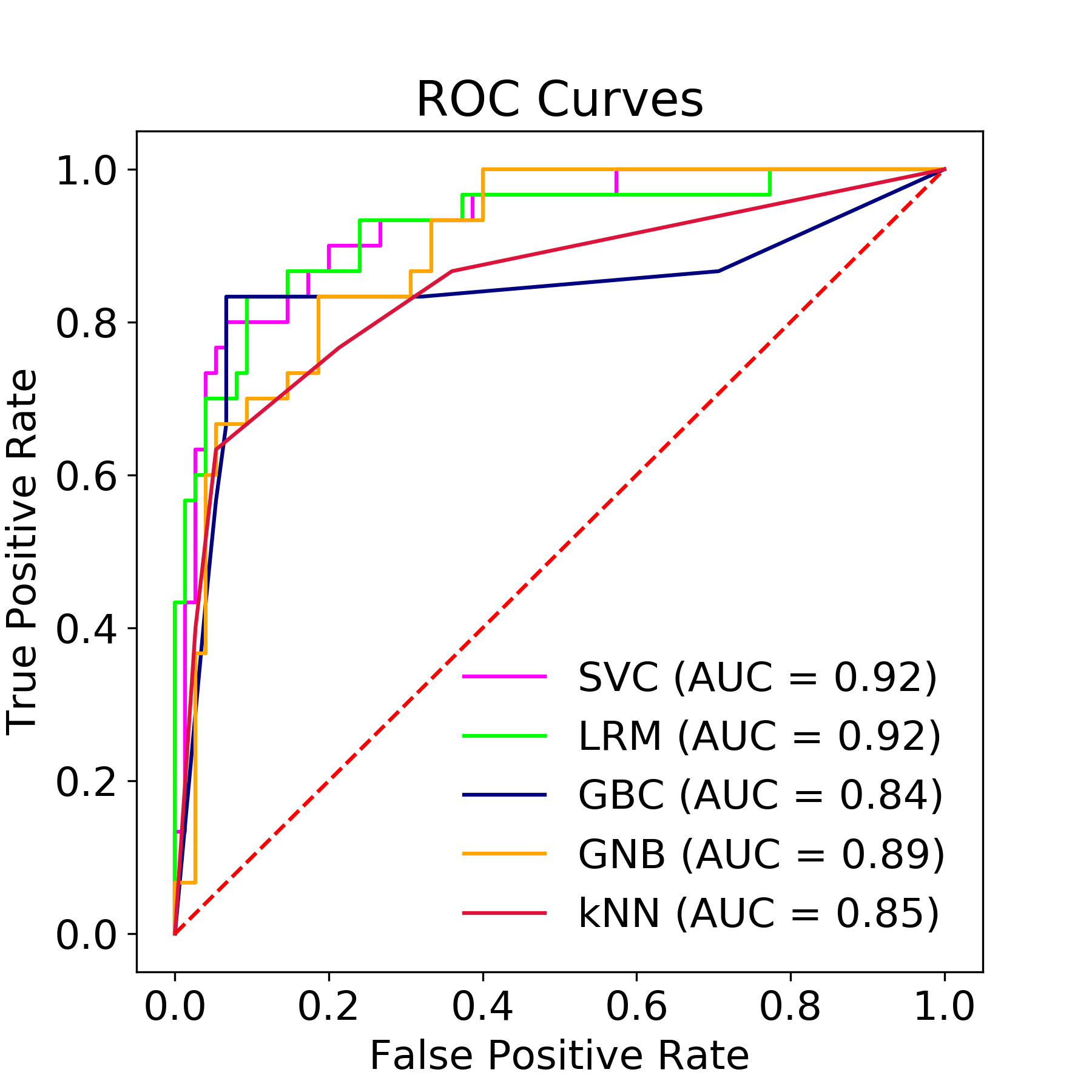} }}
    \subfloat[Si]{{\includegraphics[width=0.33\textwidth]{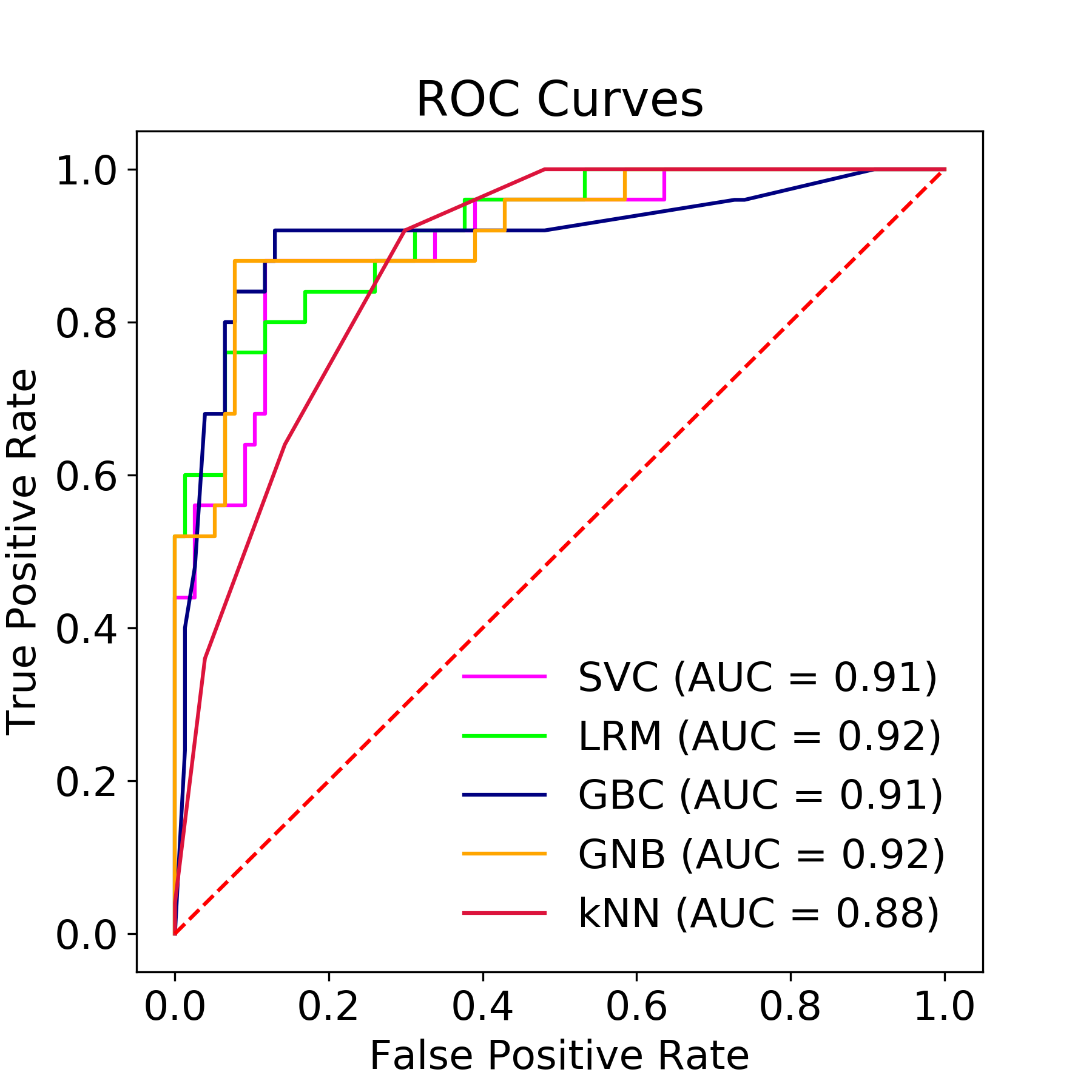} }}%
    \caption{Performance assessment of several predictive models on experimental data for all studied materials.}%
    \label{pred:model:comp:fig}%
\end{figure*}

Fig. \ref{pred:model:comp:fig} presents the ROC curves as well as the average AUC values on experimental data for the optimized predictive models for the three materials tested. Given the superior performance of the constructed features, all training and evaluation have been performed on the extended linear+quadratic+cubic feature set. For each material, there are at least two models that achieve AUC above 0.9 implying that the classification tasks are successfully learned.
The comparison between predictive models reveals that the LR model (green lines) has the best performance in terms of AUC in two out of three test materials. Moreover, SVM classifier (SVC) and LR model are the two predictive models with AUC above 0.9 in all materials. Interestingly, the high-capacity GBC performed worse than GNB. Two potential reasons for such behavior are the sample size of the datasets is not adequate for robust learning of GBC as well as GBC is sensitive to hyper-parameter tuning.

These is also a clear separation between materials in terms of how accurate the predictive models are. All predictive models for Ti6Al4V produce average AUC values that are above 0.95 while for the other two materials (i.e., Si and Steel 1.7131) the average AUC is below 0.92 yet above 0.84 for all predictive models.
Overall, the ability of the tested predictive models to accurately learn the mapping between the laser parameters and the observed structure is high. Interestingly, simpler models perform better in terms of AUC mainly due to the small number of experimental data for which simpler models generalize in a more robust manner. We explore in more details the effect of sample size in the next section.

\subsection{Effect of sample size on predictive models}

\begin{figure*}%[H]
    \centering
    \subfloat[Linear features]{{\includegraphics[width=0.4\textwidth]{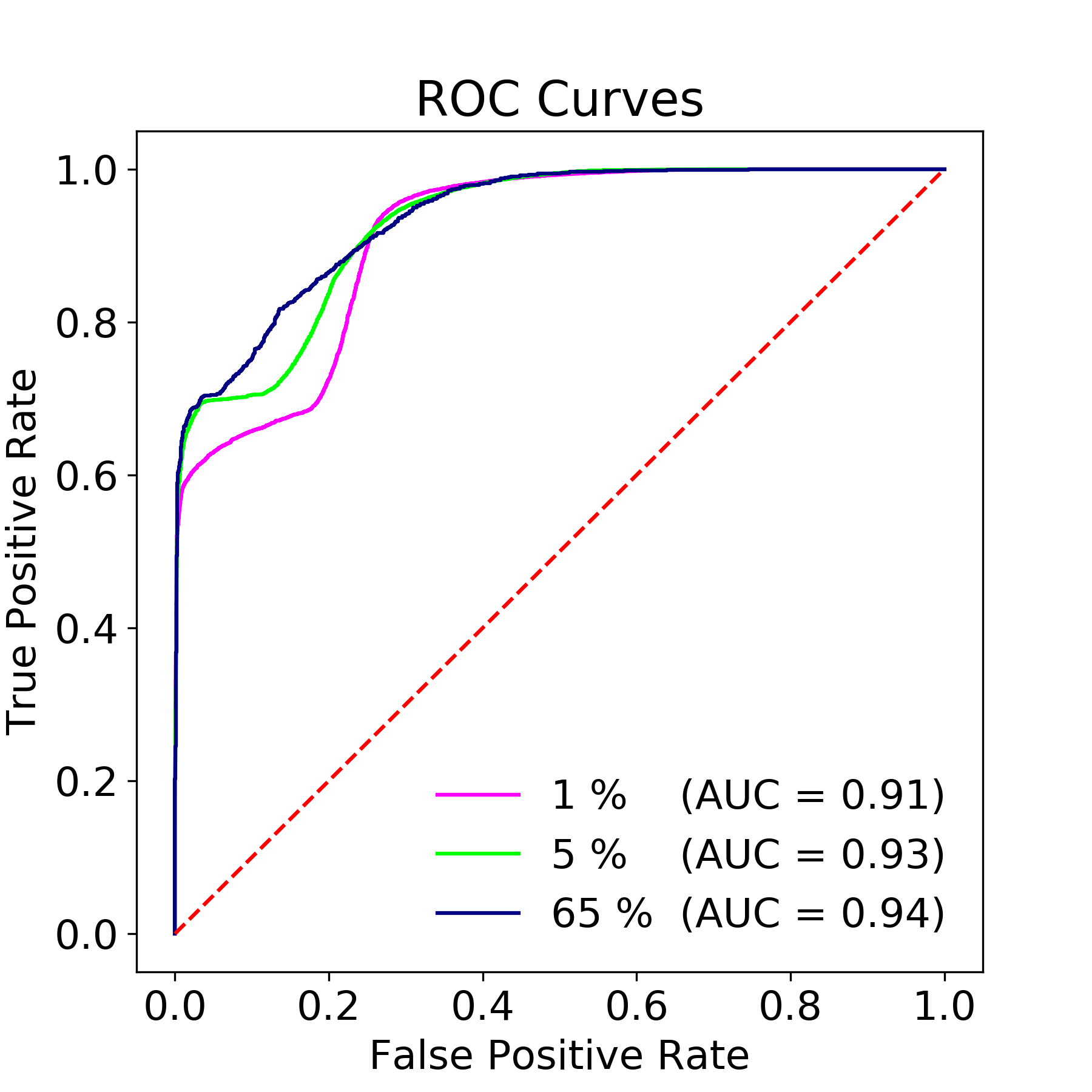} }}%
    \qquad
    \subfloat[L+Q+C features]{{\includegraphics[width=0.4\textwidth]{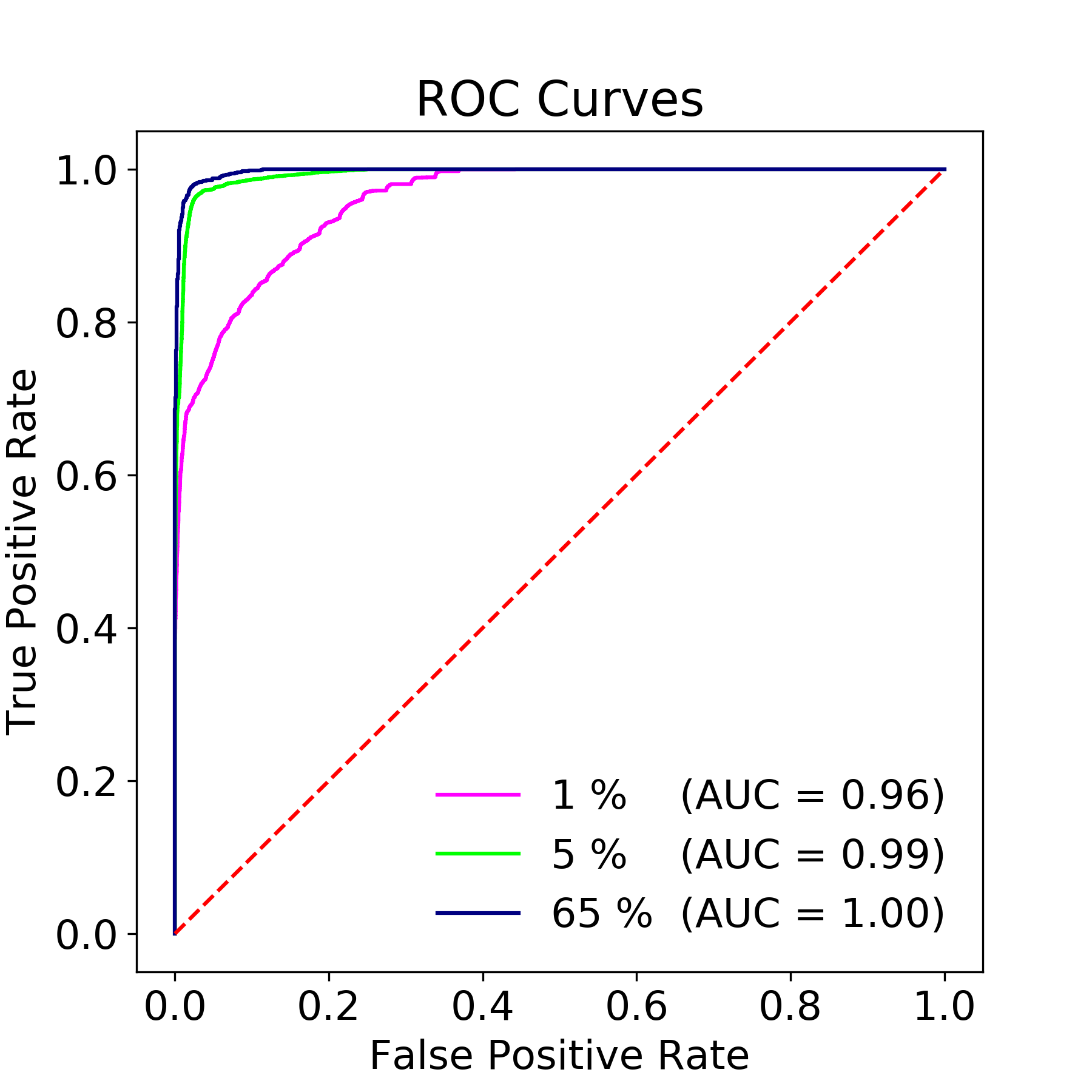} }}%
    \caption{Performance measures using the LR model and a fraction of simulated data.}%
    \label{var:sample:size:fig}%
\end{figure*}

We examine whether or not the observed performance difference between experimental and simulated datasets in Fig. \ref{various:feat:fig} can be attributed to the sample size of each dataset. Since the transition from one structure to another occurs rapidly, it is beneficial to collect as much samples as possible from the boundary regions and the resolution of the sampling may play an important role in accurately determining the boundaries. Thus, we quantify the effect of dataset size in the performance of predictive models and particularly of logistic regression model.
Fig. \ref{var:sample:size:fig} shows the ROC curves for linear features (left panel), linear+quadratic+cubic features (right panel) and various sample size percentages. In both cases, the AUC is increased as a function of the number of data instances.
%Interestingly, the increase in performance is marginal for the linear features showing that linear boundaries can be efficiently estimated with relatively fewer data. In contrast, the performance increase is substantial for the cubic features implying that more data points are required in order to finely determine the non-linear boundary.

\begin{figure*}
    \centering
    \subfloat[SVC and LR model]{{\includegraphics[width=0.4\textwidth]{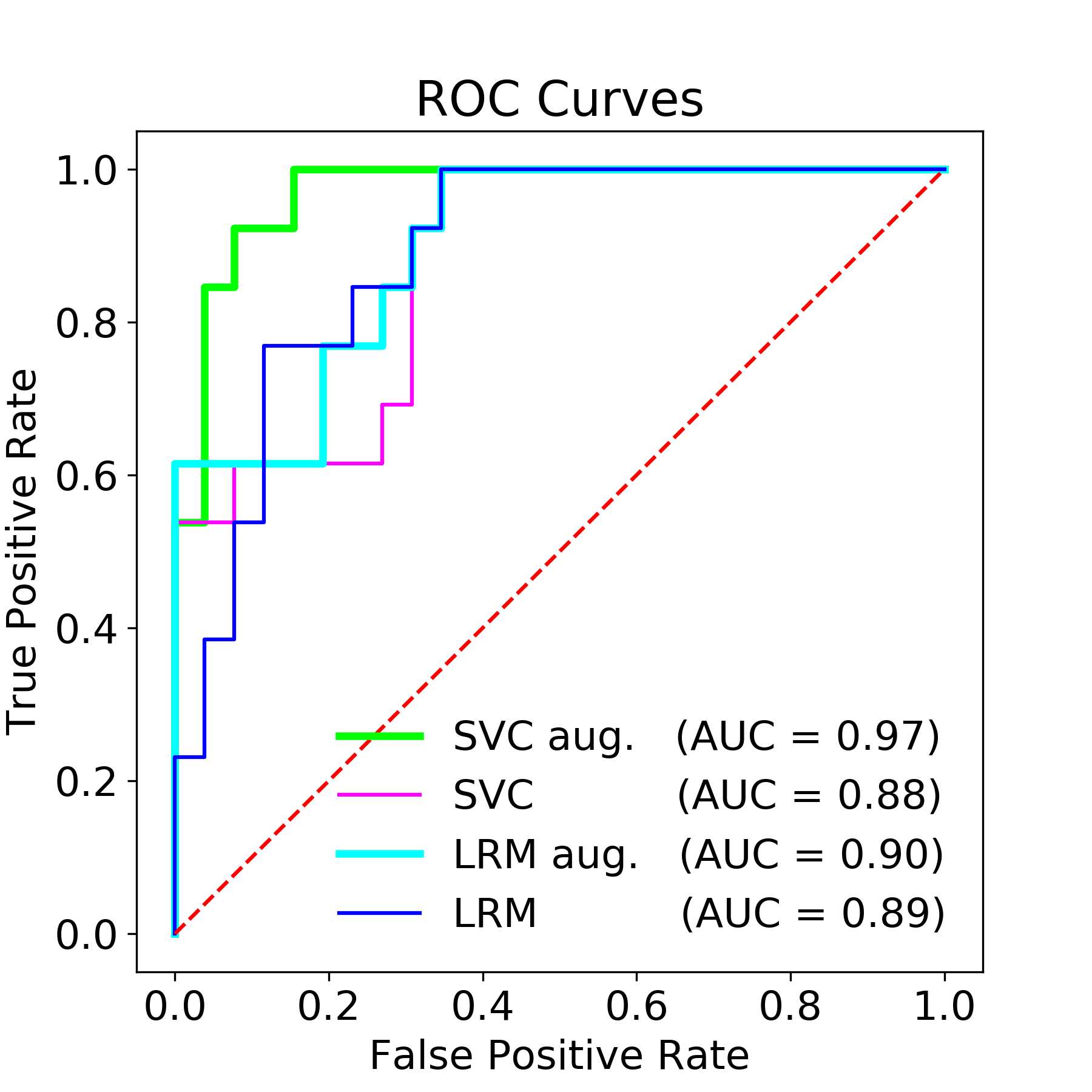} }}%
    \qquad
    \subfloat[GB and kNN classifiers]{{\includegraphics[width=0.4\textwidth]{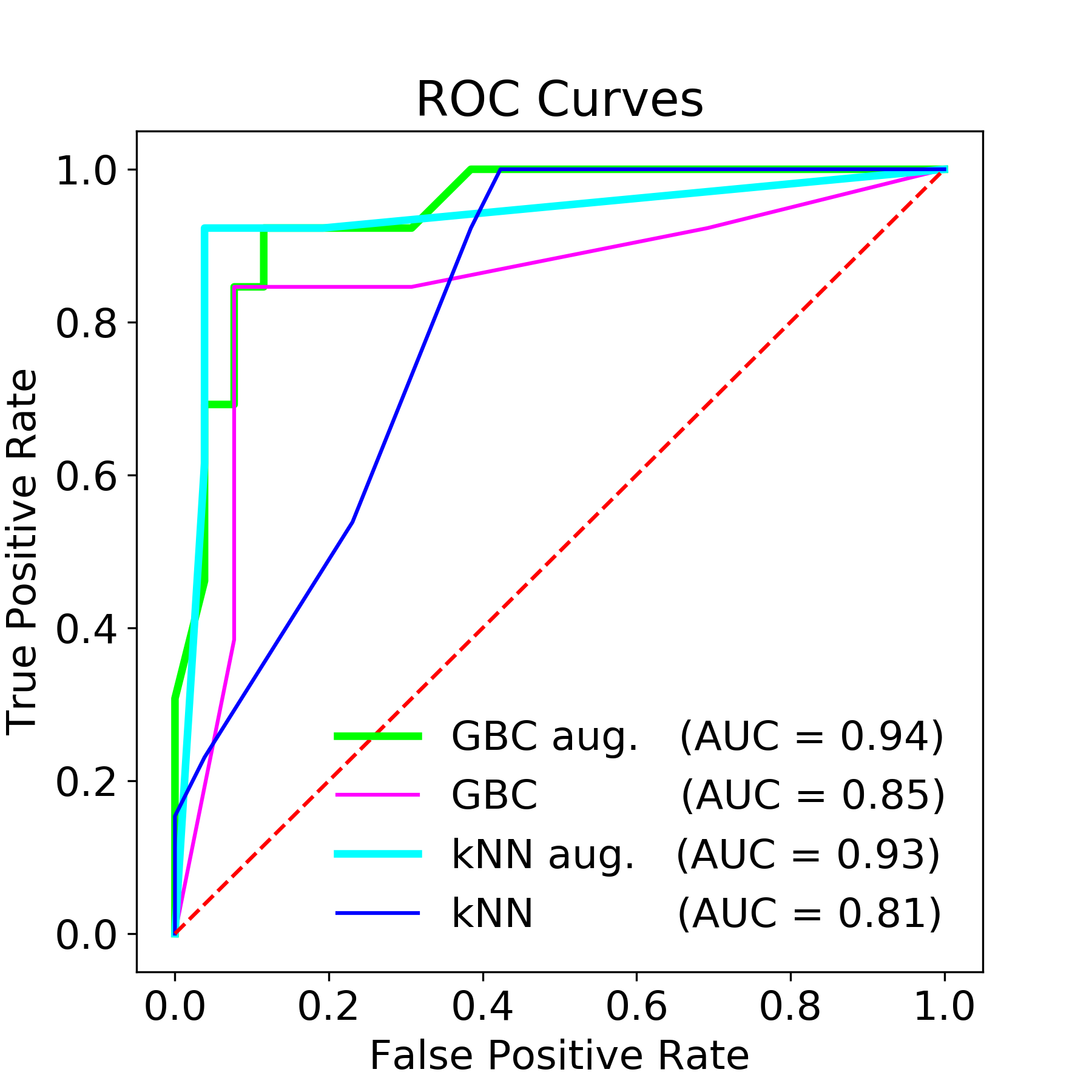} }}%
    \caption{Predictive models' performance comparison when simulated data are augmented to the experimental data for Ti6Al4V.}%
    \label{augmented:data:fig}%
\end{figure*}

Given the limited amount of experimental data, it is desirable to use the simulated data and increase the predictive performance. Unfortunately, the direct augmentation of simulated data to the experimental data will not improve the results due to the fact that they are not perfectly aligned. As explained earlier, the reason for the misalignment is that not all physical phenomena are taken into account by the materials modelling. The proposed data-driven solution is the introduction of an affine transformation which adjusts the domain of the simulated data to the domain of the experimental data.  The parameters of the affine transformation are estimated with the Cognitive-based Adaptive Optimization (CAO) algorithm
\cite{Renzaglia2010}. CAO is a derivative-free stochastic optimization method.
Fig. \ref{augmented:data:fig} and Table \ref{augmented:data:table} present the ROC curves, the AUC values as well as the relative improvement of the augmentation with the adjusted simulated data. Evidently, all predictive models trained with linear features benefited from the augmented data with the improvement being around 10\% for four out of five models. Particularly, SVC with an average AUC value of 0.97 becomes the best performing ML model.

\begin{table}%[h]
\caption{AUC for various classifiers without and with augmentation and their relative improvement.}
\begin{ruledtabular}
\begin{tabular}{cccc}
&only exp. & exp. \& sim. aug. & improvement (\%) \\
\hline
SVC&0.88&0.97&{\bf 9.7}\\
LR &0.89&0.90&{1.1}\\
GBC&0.85&0.94&{\bf 10.1}\\
GNB&0.72&0.78&{\bf 8.0}  \\
$k$NN&0.81&0.93&{\bf 13.8}\\
\end{tabular}
\end{ruledtabular}
\label{augmented:data:table}
\end{table}

\subsection{Uncertainty Regions}
\begin{figure*}%[H]
    \centering
    \subfloat[Ti6Al4V]{{\includegraphics[width=0.33\textwidth]{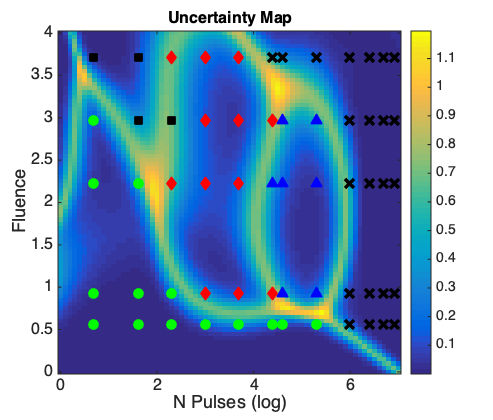} }}%    
    \subfloat[Steel (1.7131)]{{\includegraphics[width=0.33\textwidth]{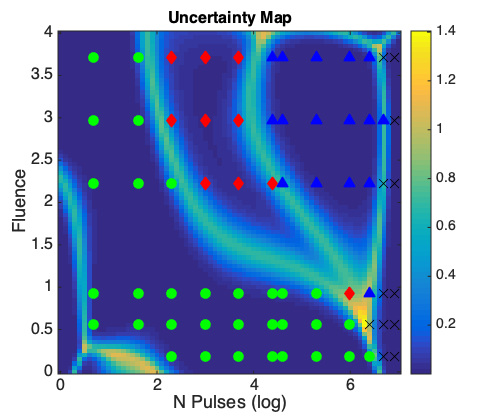} }}
    \subfloat[Si]{{\includegraphics[width=0.33\textwidth]{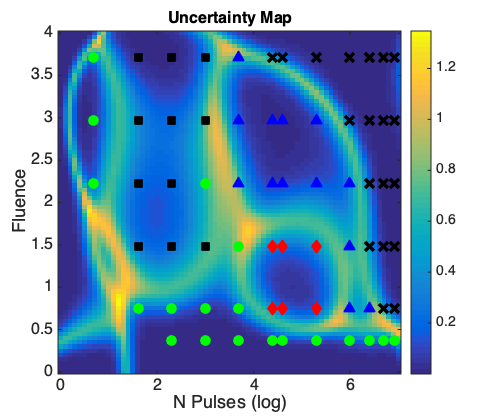} }}%
    \caption{Uncertainty quantification for the studied materials using logistic regression as predictive model.}% Stars corresponds to the higher uncertainty values as estimated by CAO.}
    \label{uq:Si:logistic:fig}
\end{figure*}

Lastly, we quantify the uncertainty of the predictive models on the parametric space where the risk of a misclassification error is high. Knowing the regions were models are not certain is helpful in designing the next experiments and extract now knowledge since the uncertain regimes contain relatively more information. On the opposite side, knowing where predictive models are confident about the structure is also useful for production purposes because clear, unequivocal structures are of practical merit. We propose to quantify the uncertainty with Shannon's information entropy function \cite{Shannon48,Cover2006} on the probability distribution generated by the predictive model. Thus, uncertainty is defined through information entropy for each point, $x$, as
\begin{equation*}
u(x) = - \sum_{c} P_{\hat{\theta}}(c|x) \log P_{\hat{\theta}}(c|x) \ ,
\end{equation*}
where $P_{\hat{\theta}}(c|x)$ is the estimated predictive distribution of the labels. Uncertainty takes the highest value when all structures have equal probabilities (i.e., uniform distribution) while the minimum value is attained when the probability for one structure is unity. Fig. \ref{uq:Si:logistic:fig} presents the uncertainty as a function of the input variables for the three studied materials. Evidently, the transition boundaries have higher uncertainty value (yellow color). In contrast, low uncertainty (dark blue color) are attained close to the center of each structure's parameter region.

% The classification errors happened at the transition boundaries from one structure to the other.

\section{Concluding Remarks and Future Directions}
\label{concl:sec}

In this work, we presented a detailed ML-based approach to estimate laser parameters for fabrication of surface patterns with fs laser beams. Departing from the traditional selection of optimal input laser parameters for a given output, usually done manually through a trial-error method, we implemented ML techniques to make calibration of laser parameters more automatic, faster and easier than the existing practices. The ML forecasting model shows very good accuracy and capability towards predicting the occurrence of all three types of self-assembled structures efficiently. The approach achieved a successful quantification of the uncertainty of each region related to particular structures and each material while automatically estimated the most uncertain points. A future extension to the model could be related to a fully automated process of structure type identification through image processing tools and structure labelling to efficiently explore the parameter space. There is no doubt that the  approach requires further validation and possibly more development, however, the predictive design is expected to transform surface patterning technique by making it more data-driven while providing routes for optimization of low-cost fabrication of products with desired properties.

Despite the impressive performance of the ML-based methodology to predict the laser parameters to produce structures with particular features, there are still several questions that arise about the analysis and the realistic application of the results. More specifically, the investigation was focused, for the sake of simplicity, on three types of structures (i.e.ripples, grooves, spikes) in the absence of regions with questionable structure type classification. Although some preliminary predictive modelling was performed to analyse those structures, a more conclusive analysis requires further investigation. 

It should finally be emphasized that the precision of the predictive modelling approach and information of critical decision points can further be enhanced by the set up of more reliable and accurate experimental protocols as well as the development of more advanced theoretical physics models. Nevertheless, the predictive model presented in this work is aimed to set the basis for a systematic fabrication methodology by reducing the number of expensive trial and error techniques and time consuming simulated experiments. Results manifested that the combination of predictive and material modelling tools are capable to reduce the time and cost required to move materials from discovery to application. Therefore, ML-based models are expected to enhance the innovation capacity of laser manufacturing companies as it constitutes a powerful tool designed for simulating and testing new techniques and methods, developing new advanced materials and products, and exploring new directions in the field of laser materials processing and manufacturing.

%\noindent \\ \\
%{\bf SUPPLEMENTARY MATERIAL}
%
%\noindent
%The Supplementary Material reports additional results on the predictive performance for various cases. We present the performance of constructed features for SVC algorithm, we provide performance results for neural networks and highlight the importance of having thousands of samples for the classification tasks at hand and we plot figures on experimental data for Si and Steel (1.7131) when augmented with simulated data.
%
%%See Supplementary Material for figures on predictive performance of constructed features for SVC algorithm. Also, for figures on Si and Steel (1.7131) where simulated data are used for the augmentation of the experimental data.

\noindent \\ \\
{\bf ACKNOWLEDGEMENTS}

\noindent
M-C.V. acknowledges financial support from $CORI$ (MIS 5031029) under Greece - Israel Call for Proposals for Joint R\&D Projects, co-financed by the national funding through the Operational Program Competitiveness, Entrepreneurship and Innovation,  and the European Regional Development Fund of the European Union. G.D.T and E.St. acknowledge funding from  $HELLAS-CH$ project (MIS 5002735), implemented under the ``Action for Strengthening Research 
and Innovation Infrastructures'' funded by the Operational Programme ``Competitiveness, Entrepreneurship and Innovation'' and co-financed by Greece and the EU (European Regional Development Fund). G.D.T, A.M, E.Sk. and E.St.  acknowledge support by the European Union's Horizon 2020 research and innovation program through the project $BioCombs4Nanofibres$ (grant agreement No. 862016). 
Y.P. acknowledges partial support by the project ``Innovative Actions in Environmental Research and Development (PErAn)'' (MIS 5002358) funded by the Operational Programme ``Competitiveness, Entrepreneurship and Innovation'' (NSRF 2014-2020). 

\noindent \\ \\
{\bf DATA AVAILABILITY}

The data that support the findings of this study are available within the article.

\noindent \\ \\
{\bf REFERENCES}

\bibliographystyle{aip}
\bibliography{aipsamp}% Produces the bibliography via BibTeX.

\appendix

\section{Predictive performance of SVC on constructed features}
\begin{figure*}[h]
    \centering
    \subfloat[Results on experimental data]{{\includegraphics[width=0.4\textwidth]{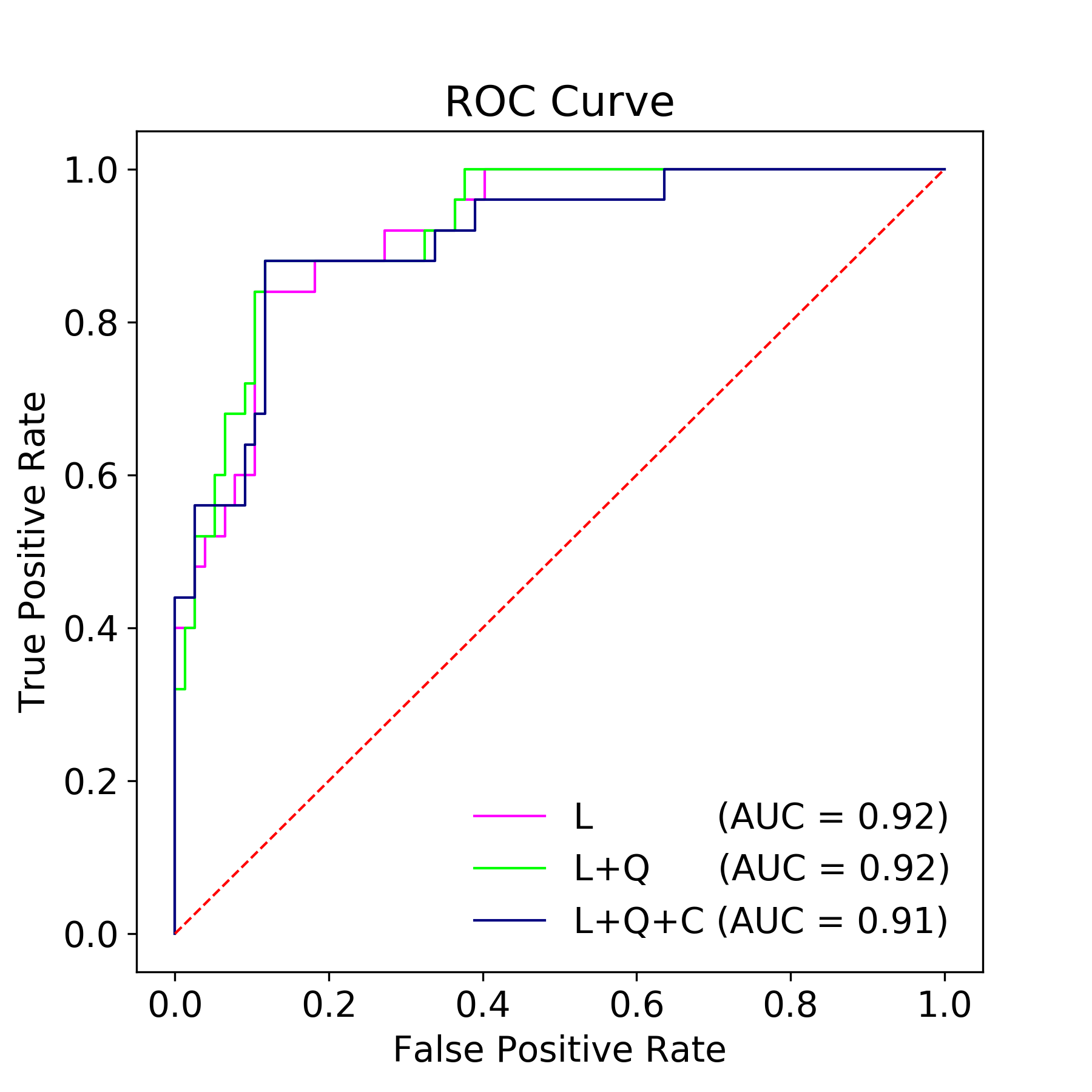} }}%
    \qquad
    \subfloat[Results on simulated data]{{\includegraphics[width=0.4\textwidth]{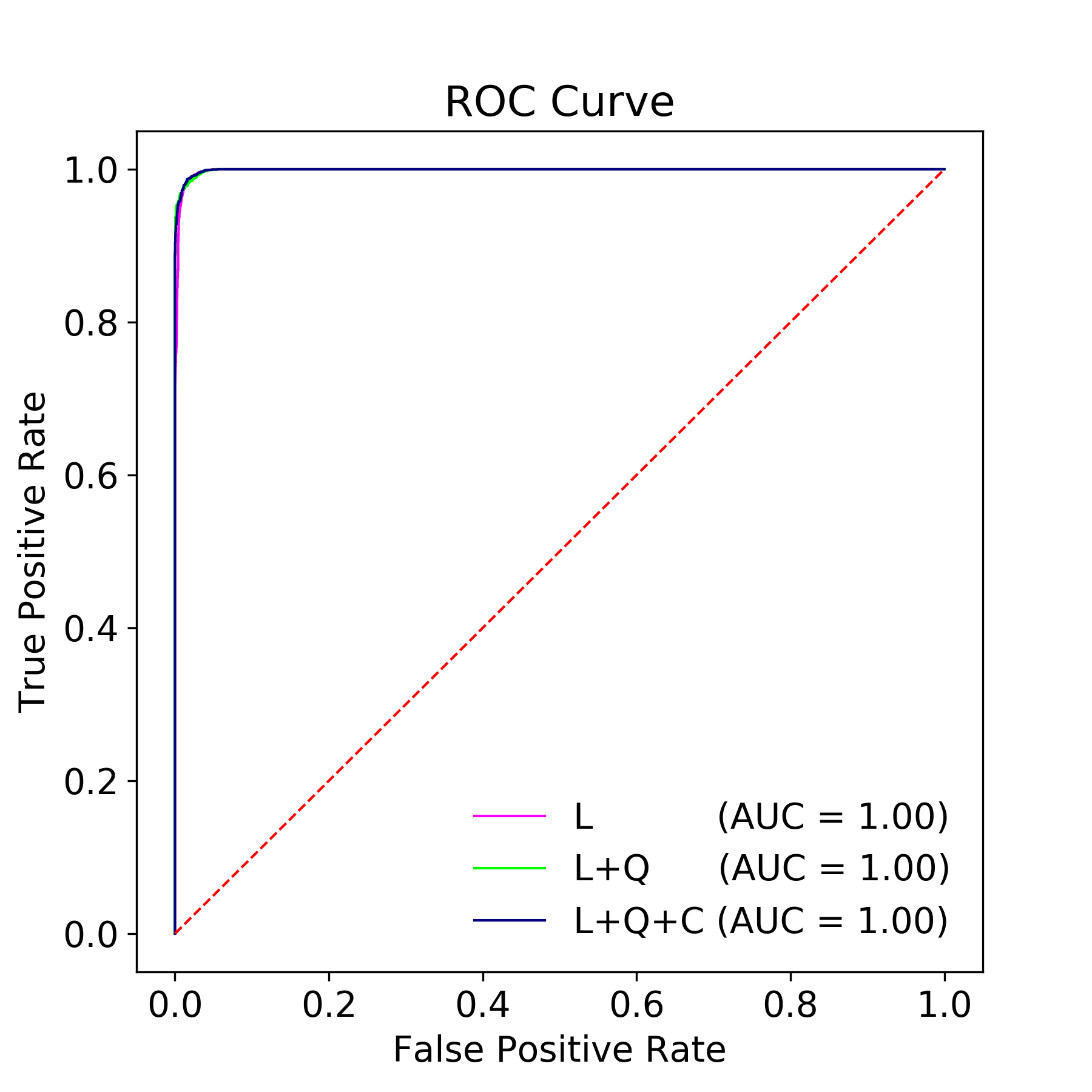} }}%
    \caption{Performance assessment of constructed features for Si. The use of quadratic or even quadratic+cubic features do not affect the accuracy of the SVC model.}%
    \label{SVC:LQC:fig}%
\end{figure*}

Fig. \ref{SVC:LQC:fig} illustrates the effect of constructed features on the predictive performance using the SVC model. It is evident that SVC did not benefit from the additional features and the performance is more or less unchanged. This is partially explained by the fact that SVC is a non-linear model capable of intrinsically constructing its own features.

%Here we highlight, that in all other algorithms the construction of quadratic and cubic features contributed significantly to the improvement of our results, but the SVC algorithm was not affected.

\section{Predictive performance of augmented data on Si and Steel}
\begin{figure*}[h]
    \centering
    \subfloat[SVC and LR model]{{\includegraphics[width=0.4\textwidth]{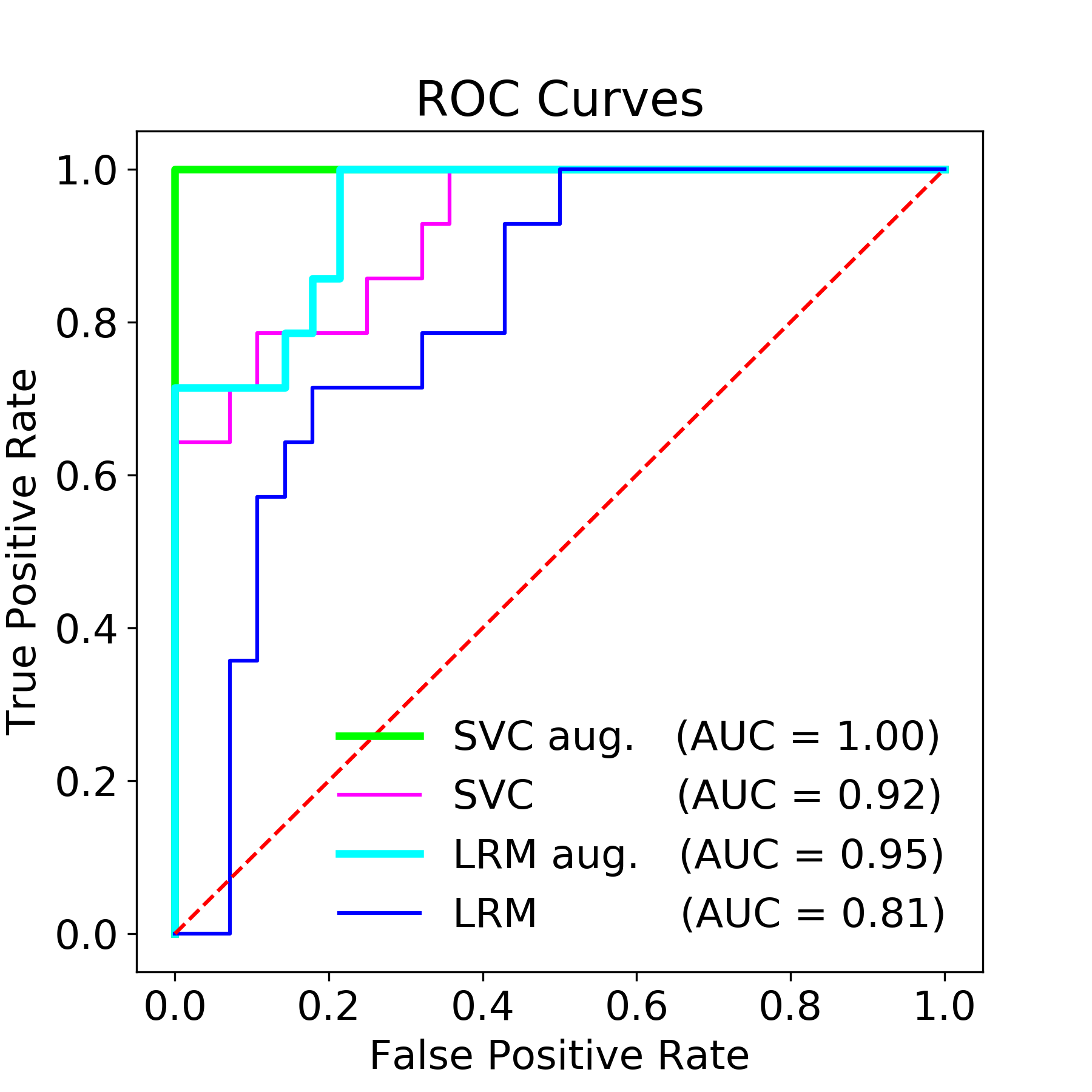} }}%
    \qquad
    \subfloat[GB and kNN classifiers]{{\includegraphics[width=0.4\textwidth]{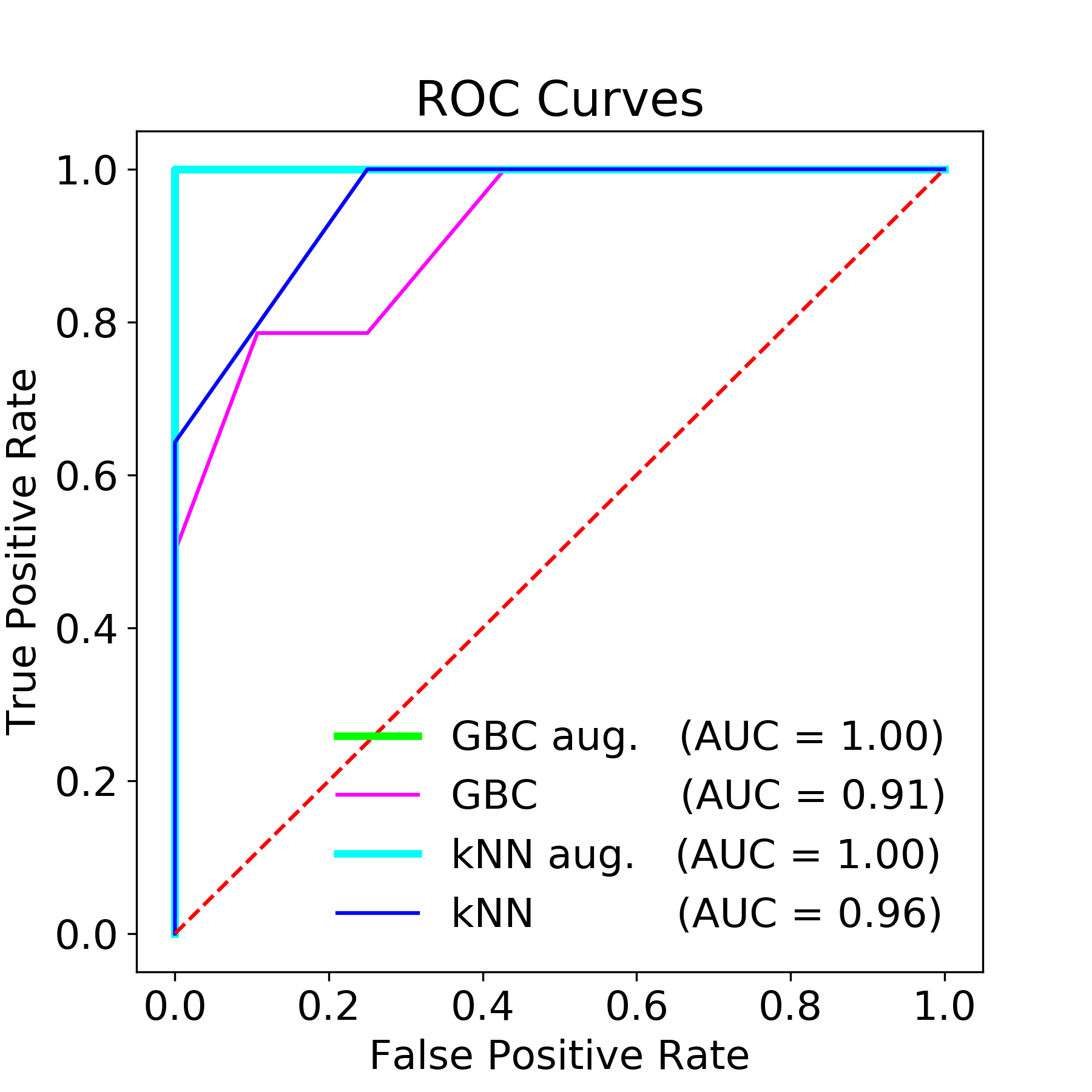} }}%
    \caption{Predictive models' performance comparison when simulated data are augmented to the experimental data for Si using linear features.}%
    \label{augmented:data:Si:fig}%
\end{figure*}

\begin{figure*}[h]
    \centering
    \subfloat[SVC and LR model]{{\includegraphics[width=0.4\textwidth]{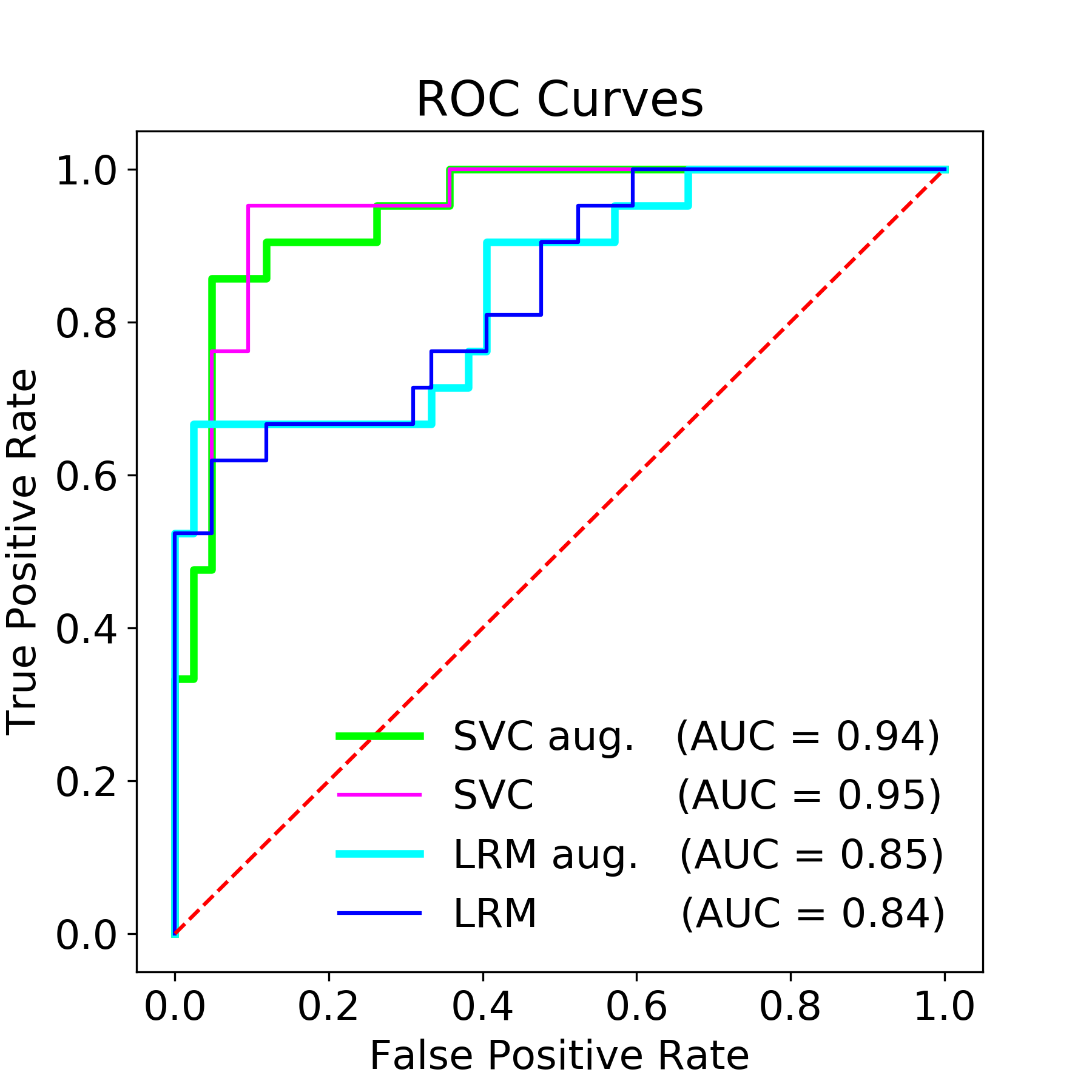} }}%
    \qquad
    \subfloat[GB and kNN classifiers]{{\includegraphics[width=0.4\textwidth]{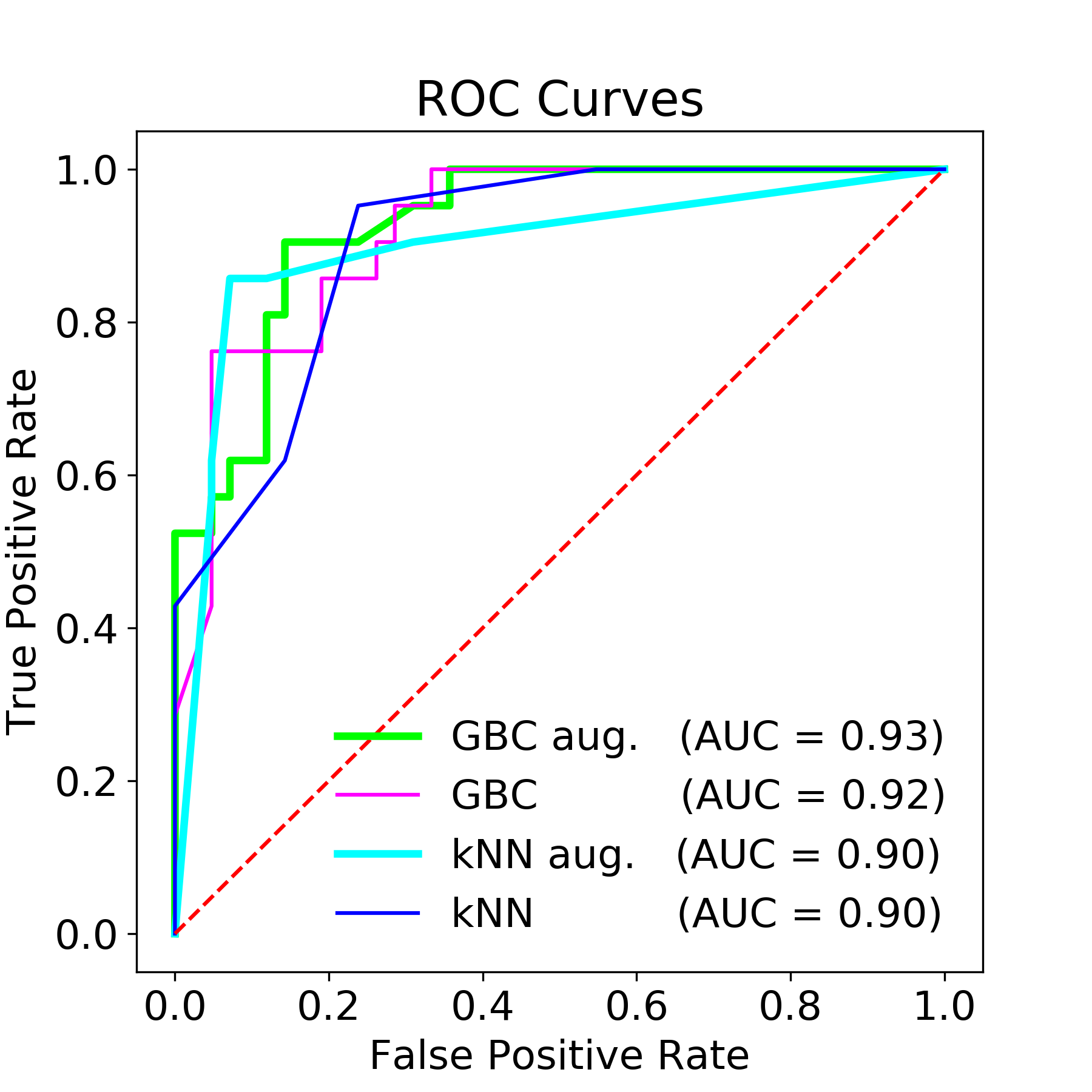} }}%
    \caption{Predictive models' performance comparison when simulated data are augmented to the experimental data for Steel (1.7131) using linear features.}%
    \label{augmented:data:St:fig}%
\end{figure*}

In this section we present the ROC curves for Si (Fig. \ref{augmented:data:Si:fig}) and Steel (1.7131) (Fig. \ref{augmented:data:St:fig}). Results for Si show that, the augmentation of experimental data with simulated ones improve considerably the predictions, as in Ti while the performance is unchanged for Steel (1.7131).
%One could notice that the results before augmentation are not as good as in fig 7 of the main article. The reason for that is that in Fig. 7 we have 5 classes of around 70 data points (the extra classes being “roughness” and “no structure”), but for the rescale we only kept the data that correspond to ripples grooves and spikes that are half of them if not less.
Tables \ref{augmented:data:Si:table} and \ref{augmented:data:St:table} quantify the improvement of the predictive models seen in the figures for Si and Steel (1.7131), respectively. Evidently, the performance improvement of the augmented dataset for a material depends on the success and accuracy of the domain transformation.
% performed, meaning how well the experimental data coincide with the simulated data. 

% not significant for Steel and in some cases the outcome is worse. In contrast, for Si we have improvement for all the algorithms tested.

\begin{table}%[h]
\caption{AUC for various classifiers without and with augmentation and their relative improvement for Si.}
\begin{ruledtabular}
\begin{tabular}{cccc}
&only exp. & exp. \& sim. aug. & improvement (\%) \\
\hline
SVC&0.92&1&{\bf8.3 }\\
LR &0.81&0.95&{\bf15.9}\\
GBC&0.91&1&{\bf 9.4}\\
GNB&0.86&0.91&{\bf 5.6}  \\
$k$NN&0.96&1&{\bf 4.1}\\
\end{tabular}
\end{ruledtabular}
\label{augmented:data:Si:table}
\end{table}

\begin{table}%[h]
\caption{AUC for various classifiers without and with augmentation and their relative improvement for Steel (1.7131).}
\begin{ruledtabular}
\begin{tabular}{cccc}
&only exp. & exp. \& sim. aug. & improvement (\%) \\
\hline
SVC&0.95&0.94&{\bf-1.1 }\\
LR &0.84&0.85&{1.2}\\
GBC&0.92&0.93&{ 1.1}\\
GNB&0.85&0.63&{\bf -29.7}  \\
$k$NN&0.90&0.90&{ 0}\\
\end{tabular}
\end{ruledtabular}
\label{augmented:data:St:table}
\end{table}

\section{Predictive performance with neural networks}
\begin{figure*}[h]
    \centering
    \subfloat[Linear features]{{\includegraphics[width=0.4\textwidth]{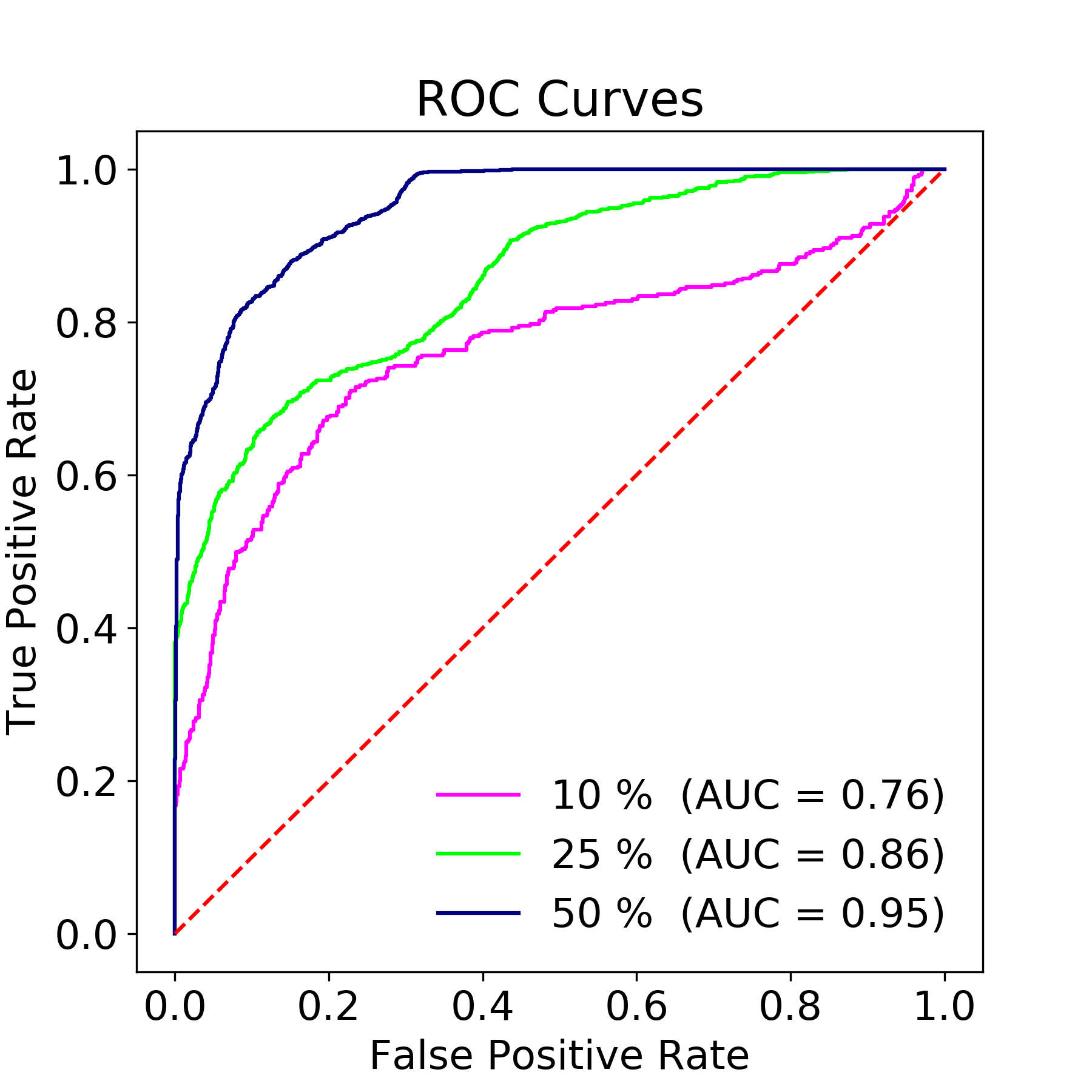} }}%
    \qquad
    \subfloat[L+Q+C features]{{\includegraphics[width=0.4\textwidth]{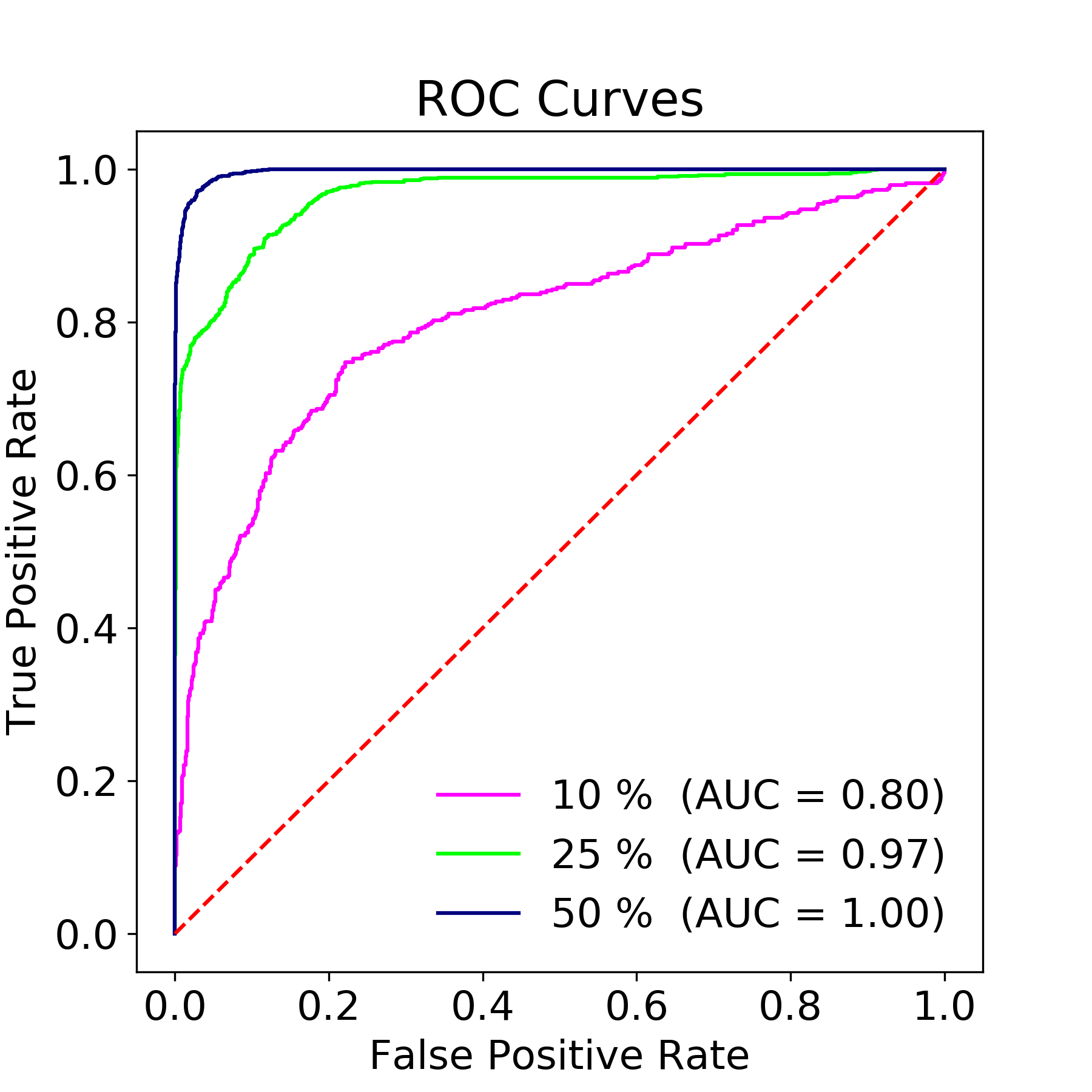} }}%
    \caption{Predictive performance results using a neural network with 1 hidden layer with 50 units on simulated data.}%
    \label{NN1:fig}%
\end{figure*}

\begin{figure*}[h]
    \centering
    \subfloat[Linear features]{{\includegraphics[width=0.4\textwidth]{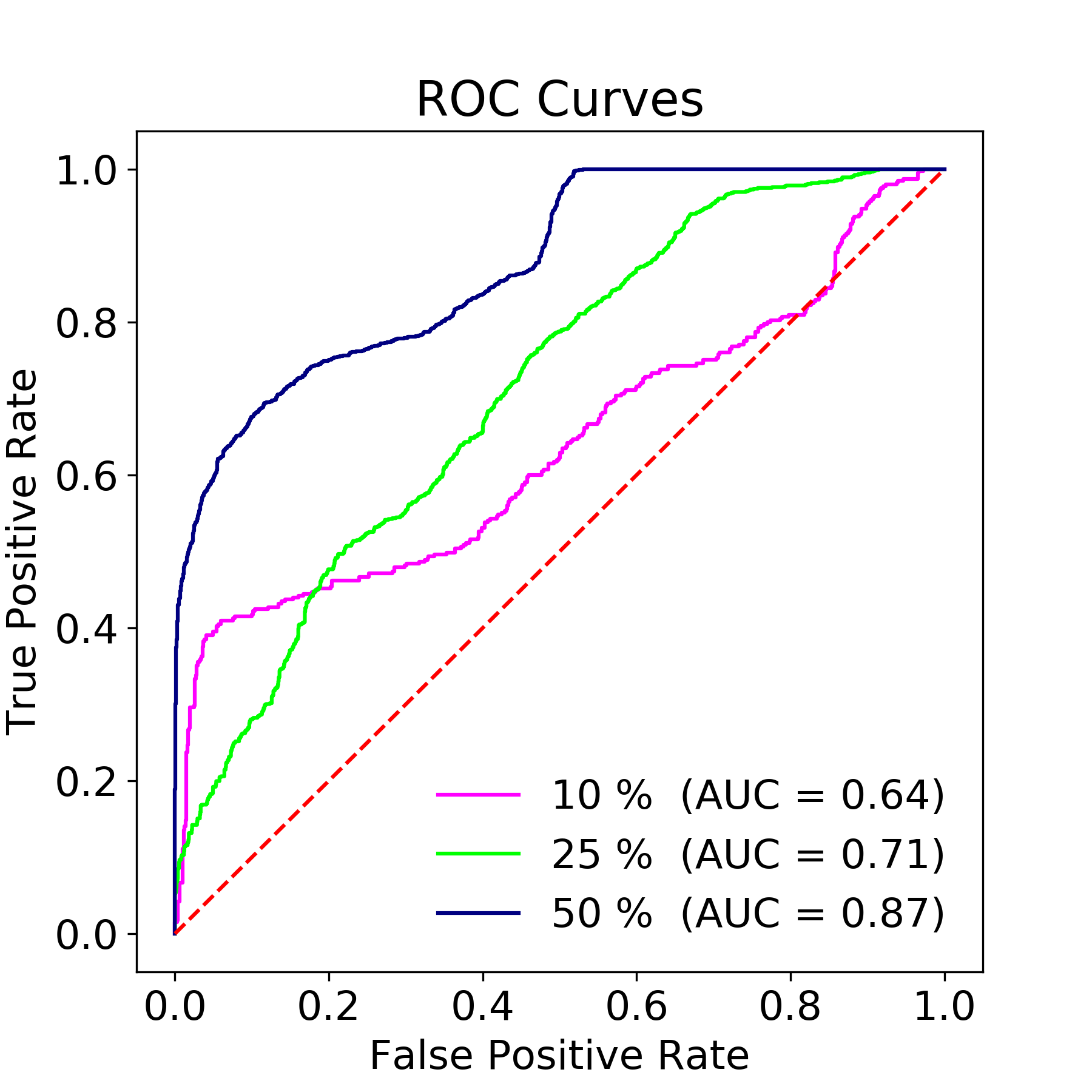} }}%
    \qquad
    \subfloat[L+Q+C features]{{\includegraphics[width=0.4\textwidth]{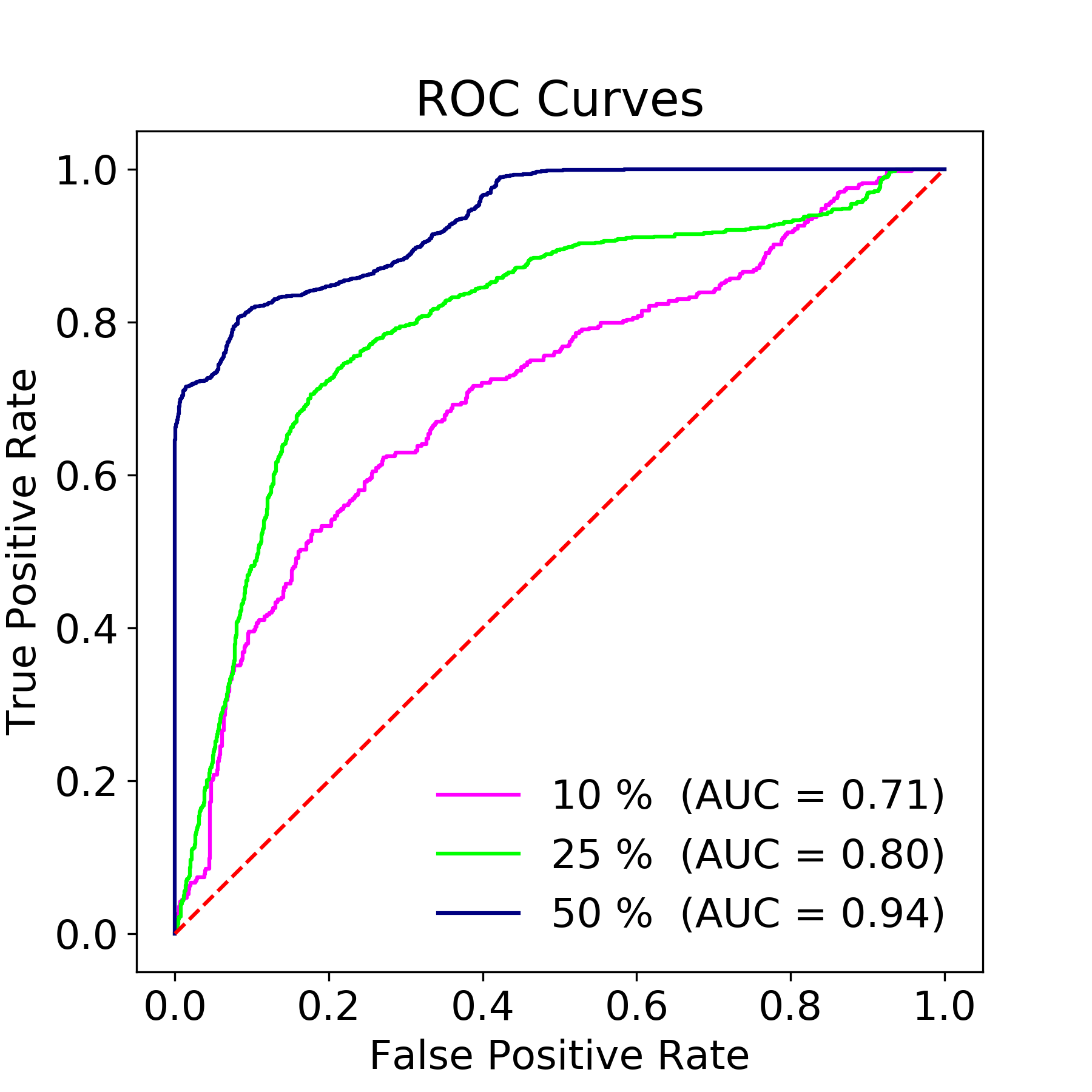} }}%
    \caption{Predictive performance results using a neural network with 1 hidden layer with 10 units  on simulated data.}%
    \label{NN2:fig}%
\end{figure*}

\begin{figure*}[h]
    \centering
    \subfloat[1 hidden layer with 10 units]{{\includegraphics[width=0.4\textwidth]{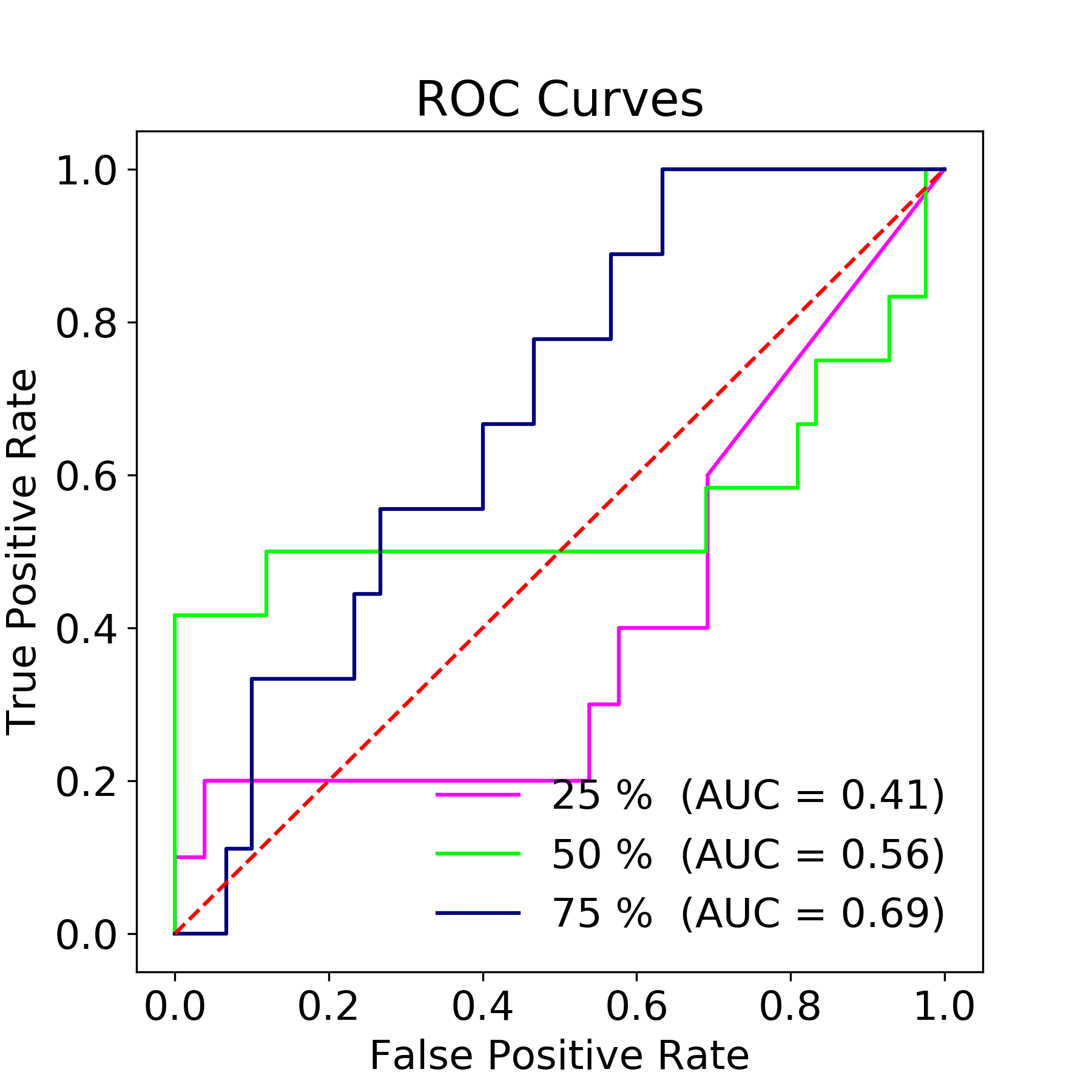} }}%
    \qquad
    \subfloat[1 hidden layer with 50 units]{{\includegraphics[width=0.4\textwidth]{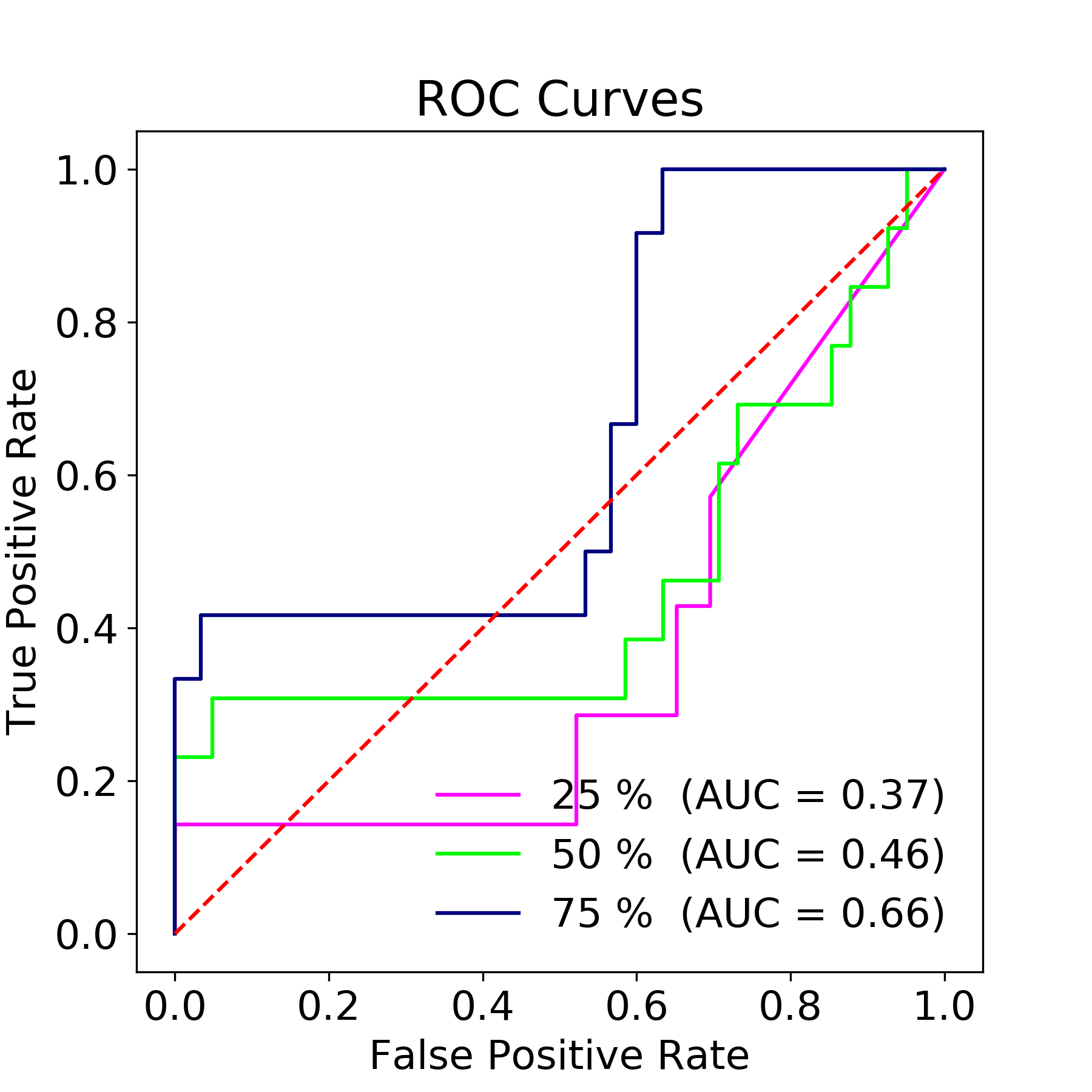} }}%
    \caption{Predictive performance results using a neural network using linear+quadratic+cubic features on experimental data.}%
    \label{NNexp:fig}%
\end{figure*}

In this section we present the predictive performance of neural networks. We first demonstrate the performance of two neural networks with one hidden layer and 50 units (Fig \ref{NN1:fig}) and 10 units (Fig \ref{NN2:fig}) on simulated data. We observe that the neural network with more units enjoy higher AUC. Moreover, the constructed features assisted the performance of the neural networks. We also remark that the number of samples used for the training has a significant effect on the performance which deteriorates considerably.
The deterioration is even higher when the training is performed on experimental data as shown in Fig. \ref{NNexp:fig}. The lack of sufficient experimental data makes the training of neural networks infeasible.

%Here we illustrate why Neural Networks were not used in our main study. Due to the lack of experimental data a NN algorithm  is meant to fail (fig. \ref{NNexp:fig}).

%In this section we present the predictive performance of neural networks. can be very good, with a high number of samples and once more the use of all linear+quadratic+cubic features (fig. \ref{NN1:fig} \& \ref{NN2:fig}).

\section{Uncertainty regions for other predictive models}
\begin{figure*}%[H]
    \centering
    \subfloat[Ti6Al4V]{{\includegraphics[width=0.33\textwidth]{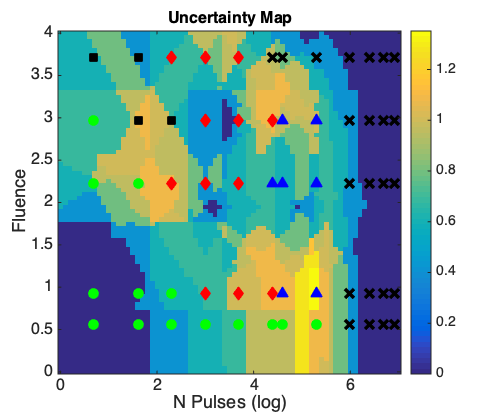} }}%    
    \subfloat[Steel (1.7131)]{{\includegraphics[width=0.33\textwidth]{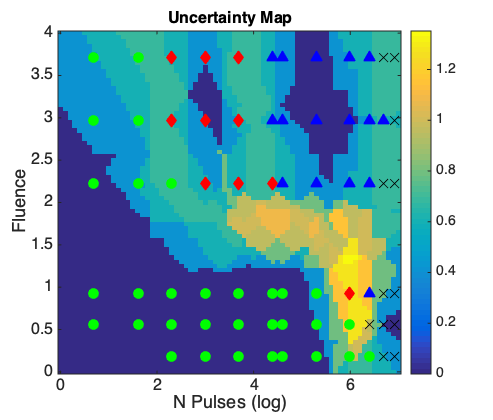} }}
    \subfloat[Si]{{\includegraphics[width=0.33\textwidth]{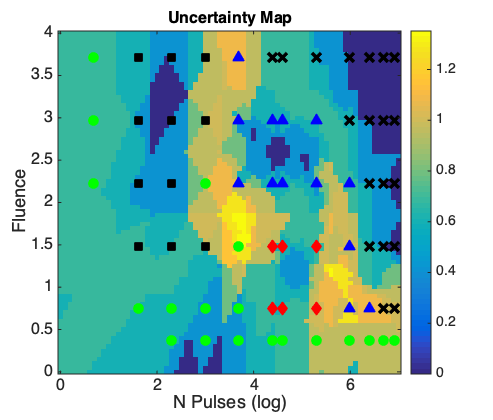} }}%
    \caption{Uncertainty quantification for the studied materials using k-NN as predictive model.}
    \label{uq:knn:fig}
\end{figure*}

\begin{figure*}%[H]
    \centering
    \subfloat[Ti6Al4V]{{\includegraphics[width=0.33\textwidth]{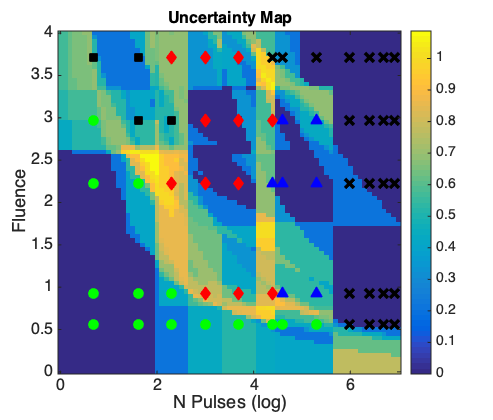} }}%    
    \subfloat[Steel (1.7131)]{{\includegraphics[width=0.33\textwidth]{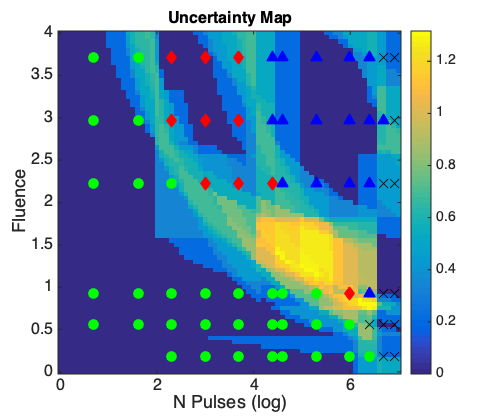} }}
    \subfloat[Si]{{\includegraphics[width=0.33\textwidth]{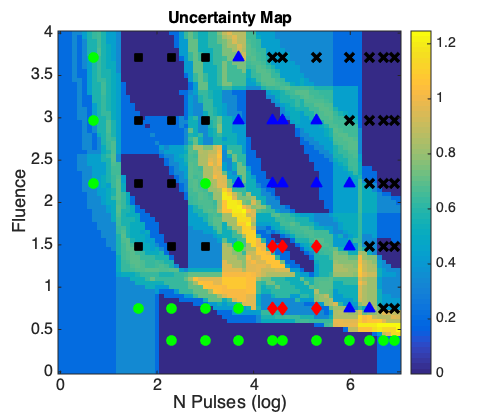} }}%
    \caption{Uncertainty quantification for the studied materials using Random Forest as predictive model.}
    \label{uq:rf:fig}
\end{figure*}

Figs. \ref{uq:knn:fig} and \ref{uq:rf:fig} present the estimated uncertainty using k-NN and random forest, respectively. Both predictive models are trained on the experimental data. Uncertainty regions are qualitatively similar. Nevertheless, uncertainty estimate using random forest are more focused around the transition boundaries.

\end{document}